\def\lesssim{\mathrel{\hbox{\rlap{\hbox{\lower3pt\hbox{${\sim}$}}}\hbox{\raise2pt\hbox{$<$}}}}}
\documentclass[useAMS,usenatbib]{mn2e}
\usepackage{cleveref}
\usepackage{natbib}
\usepackage{graphicx}
\usepackage{amsmath}
\usepackage{epstopdf}
\usepackage{subcaption}
\usepackage{float}

\newcommand{\Myr}{\,\mbox{Myr}}

\def\lesssim{\mathrel{\hbox{\rlap{\hbox{\lower3pt\hbox{$\sim$}}}\hbox{\raise2pt\hbox{$<$}}}}}
\def\gtrsim{\mathrel{\hbox{\rlap{\hbox{\lower3pt\hbox{$\sim$}}}\hbox{\raise2pt\hbox{$>$}}}}}
\def\gtreq{\mathrel{\hbox{\rlap{\hbox{\lower3pt\hbox{$-$}}}\hbox{\raise2pt\hbox{$>$}}}}}
\def\lesssim{\mathrel{\hbox{\rlap{\hbox{\lower3pt\hbox{${\sim}$}}}\hbox{\raise2pt\hbox{$<$}}}}}

\title[Multiplicity of disc-bearing stars in US and UCL]{Multiplicity of disc-bearing stars in Upper Scorpius and Upper Centaurus-Lupus}
\author[R. L. Kuruwita et al.]{Rajika L. Kuruwita$^{1}$, Michael Ireland$^{1}$, Aaron Rizzuto$^{2}$, Joao Bento$^{1}$, \newauthor Christoph Federrath$^{1}$\\ $^{1}$Research School of Astronomy and Astrophysics, Australian National University (ANU), Canberra, ACT 2611, Australia\\ $^{2}$Department of Astronomy, The University of Texas at Austin, Austin, TX 78712, USA}
\begin{document}

\pagerange{\pageref{firstpage}--\pageref{lastpage}} \pubyear{2015}

\maketitle

\label{firstpage}

\begin{abstract}
We present observations of disc-bearing stars in Upper Scorpius (US) and Upper Centaurus-Lupus (UCL) with moderate resolution spectroscopy in order to determine the influence of multiplicity on disc persistence after $\sim5-20\,\mathrm{Myr}$. Discs were identified using infra-red (IR) excess from the Wide-field Infra-red Survey Explorer (WISE) survey. Our survey consists of 55 US members and 28 UCL members, using spatial and kinematic information to assign a probability of membership. Spectra are gathered from the ANU 2.3m telescope using the Wide Field Spectrograph (WiFeS) to detect radial velocity variations that indicate the presence of a companion. We identify 2 double-lined spectroscopic binaries, both of which have strong IR excess. We find the binary fraction of disc-bearing stars in US and UCL for periods up to 20 years to be $0.06^{0.07}_{0.02}$ and $0.13^{0.06}_{0.03}$ respectively. Based on the multiplicity of field stars, we obtain an expected binary fraction of $\sim0.12^{0.02}_{0.01}$. The determined binary fractions for disc-bearing stars does not vary significantly from the field, suggesting that the overall lifetime of discs may not differ between single and binary star systems. 
\end{abstract}


\section{Introduction}
\label{sec:introduction}

In the age of Kepler and other planet finding surveys a number of planets have been discovered around binary star systems. Some of these planets are in S-Type orbits where the planet orbits one star in the binary system (e.g. $\gamma$ Cephei Ab \citep{neuhauser_direct_2007} and HD 196885 Ab \citep{chauvin_characterization_2007}) and others are in P-Type, or circumbinary orbits, where the planet orbits both stars (e.g. Kepler-47b and c \citep{orosz_kepler-47:_2012}, PH-1 \citep{schwamb_planet_2013} and ROXs 42Bb \citep{kraus_three_2014}). When considering planet formation, current models mostly focus on single star systems. This picture is, however, insufficient, given that a large fraction of stars are in binary star systems \citep{raghavan_survey_2010} and therefore a complete understanding of planet formation must take these environments into account.

Current planet occurrence rates around binary stars show that the frequency drops for systems with separations $a\sim40\,\mathrm{AU}$ \citep{kraus_impact_2016}, which might indicate that it is difficult for planets to form around binaries of this separation. Could this be because the disc material from which the planets would form is destroyed faster in binary star systems than around single stars? There are various mechanisms that contribute to the dispersal of protoplanetary discs such as accretion, photo-evaporation and outflows. How these mechanisms change and affect the discs in binary star systems has been investigated using simulations (e.g. \citealt{artymowicz_dynamics_1994, kuruwita_binary_2017}) which find that circumstellar discs are truncated to $r_{disc} \sim a/3$ and outflows from binary star systems are less efficient at carrying away mass and momentum compared to single star counterparts.

\cite{harris_resolved_2012} and \cite{cox_protoplanetary_2017} found that circumstellar discs in binaries were smaller and fainter, suggesting that they may be dispersed faster. On the surface it seems that discs have a shorter lifetime around stars in binary systems, and as a result shorten the time in which planets may form. However, they also found that circumbinary discs have at least an order of magnitude higher millimetre flux densities compared to circumstellar discs around binaries of the same separation, suggesting they are larger and have more material in the circumbinary disc to form planets. There is also a handful of other considerably old circumbinary discs (e.g. AK Sco \citep[$18\pm1\Myr$,][]{czekala_disk-based_2015}, HD 98800 B \citep[$10\pm5\,\mathrm{Myr}$,][]{furlan_hd_2007}, V4046 Sgr \citep[$12$--$23\,\mathrm{Myr}$,][]{rapson_combined_2015} and St 34 \citep[also known as HBC 425, $\sim$25$\,\mathrm{Myr}$,][]{hartmann_accretion_2005} compared to the typical lifetime of protoplanetary discs of $3\,\mathrm{Myr}$ (\citealt{haisch_disk_2001,mamajek_initial_2009}). If circumbinary discs always have a significantly longer lifetime than discs around single stars, it would provide a greater opportunity for planets to form in these systems.

In this work we aim to determine what fraction of disc-bearing stars are in binary star systems in the OB associations Upper Scorpius (US) and Upper Centaurus-Lupus (UCL). These OB associations are part of the larger Scorpius-Centaurus-Lupus-Crux (Sco-Cen) association which is the nearest region of recent massive star formation at a distance of $\sim140$~pc \citep{zeeuw_hipparcos_1999}. The subgroups of the Sco-Cen association show a well known age gradient from $\sim5$~Myr at high galactic latitudes to $\sim26$~Myr in the galactic plane \citep{pecaut_star_2016}, making this region an ideal place to study evolution of pre-main sequence stars and protoplanetary discs. The ages of US and UCL sub-groups in the Sco-Cen association are $\sim$11$\,\mathrm{Myr}$ \citep{pecaut_revised_2012} and $\sim$17$\,\mathrm{Myr}$ \citep{mamajek_post-t_2002} respectively. These regions are expected to have $\sim10^4$ G/K/M pre-main sequence stars based on any IMF. Many studies have looked to discover and characterise this low mass population (e.g. \citealt{zeeuw_hipparcos_1999, preibisch_history_1999, mamajek_post-t_2002, rizzuto_multidimensional_2011, rizzuto_new_2015}). Previous work on the presence of discs around these members (\citealt{luhman_disk_2012, rizzuto_wise_2012, pecaut_star_2016}) show a general increase in disc fraction with later spectral types. This is consistent with previous studies showing that protoplanetary disc lifetime is short around higher mass stars \citep{ribas_protoplanetary_2015}. However, none of these works investigated the binary fraction of these disc-bearing members. Work on binary fractions in the Sco-Cen region have primarily focused on the the higher mass B/A/F stars (\citealt{kouwenhoven_primordial_2005,kouwenhoven_primordial_2007}) finding binary fractions of $>70\%$. 

In order to investigate the influence of binarity on disc lifetime, we look for radial velocity variation in selected disc-bearing stars in US and UCL to determine the binary fraction. The Upper Scorpius and Upper Centaurus-Lupus regions are relatively old considering the typical protoplanetary disc lifetime, but it is at these older ages that we believe any variation in disc fractions between single and binary stars would be amplified. \Cref{sec:method} describes identification of US and UCL members, the target selection criteria, observations and how radial velocities are determined. In \Cref{sec:resultsanddiscussion} we discuss our estimated binary fractions and Bayesian analysis. In \Cref{sec:discussion} we discuss our results and caveats of our work.

\section{Method and observations}
\label{sec:method}

\Cref{ssec:membership} and \ref{ssec:wise} describe how the targets for our survey are selected. The observations and instrument specifications are detailed in \Cref{ssec:wifes}. \Cref{ssec:ObjRVs} describes how radial velocities are obtained from the observations.

\subsection{Upper Scorpius and Upper Centaurus-Lupus membership probability}
\label{ssec:membership}

We selected candidate Upper Scorpius and Upper Centaurus-Lupus members using kinematic and photometric data from UCAC4, 2MASS, USNO-B and APASS (\citealt{zacharias_fourth_2013,skrutskie_two_2006,monet_usno-b_2003,henden_data_2012}) using the Bayesian membership selection method of \citet{rizzuto_multidimensional_2011} and \citet{rizzuto_new_2015}, which uses kinematic and spatial information to assign membership probabilities. We took the proper motions from the UCAC4 catalogue \citep{zacharias_fourth_2013} and photometry from 2MASS and APASS (\citealt{skrutskie_two_2006,henden_data_2012}, summarised in \Cref{table:magnitudes1}), and used the photometry and a 15\,Myr pre-main sequence isochrone \citep{siess_internet_2000} to estimate each candidate member’s distance. We then treated the proper-motion and estimated distance together to calculate the membership probability by comparing to both the model association velocity \citep{rizzuto_multidimensional_2011} and Galactic thin disc velocity distributions \citep{robin_synthetic_2003}. This selection was magnitude limited, and covered all stars in the UCAC4 catalogue with $10<V<16$, and comprised of several thousand candidate members with membership probability greater than 25\%. These candidates were then searched for disc-indicating IR excesses at 12 and 24\,um (see \Cref{ssec:wise}). While a purely kinematic selection on the UCAC4 proper motions and photometric distances alone would not be sufficient to assign membership to G,K and M-type stars, in combination with an infrared excess indicating the presence of a primordial or debris circumstellar disc these members can be considered robust.

\subsection{IR excess to identify discs}
\label{ssec:wise}

Discs around young stars are often identified with IR observations, as an excess emission at these wavelengths can be detected when compared with the expected stellar photosphere flux of the star alone. The wavelength of IR emission correlates with the distance from the central star/stars with shorter wavelength emission originating at smaller radii and longer wavelengths originating from larger radii. This is due to the thermal profile of the discs \citep{lada_nature_1984}.

\begin{figure}
	\centerline{\includegraphics[width=1.05\linewidth]{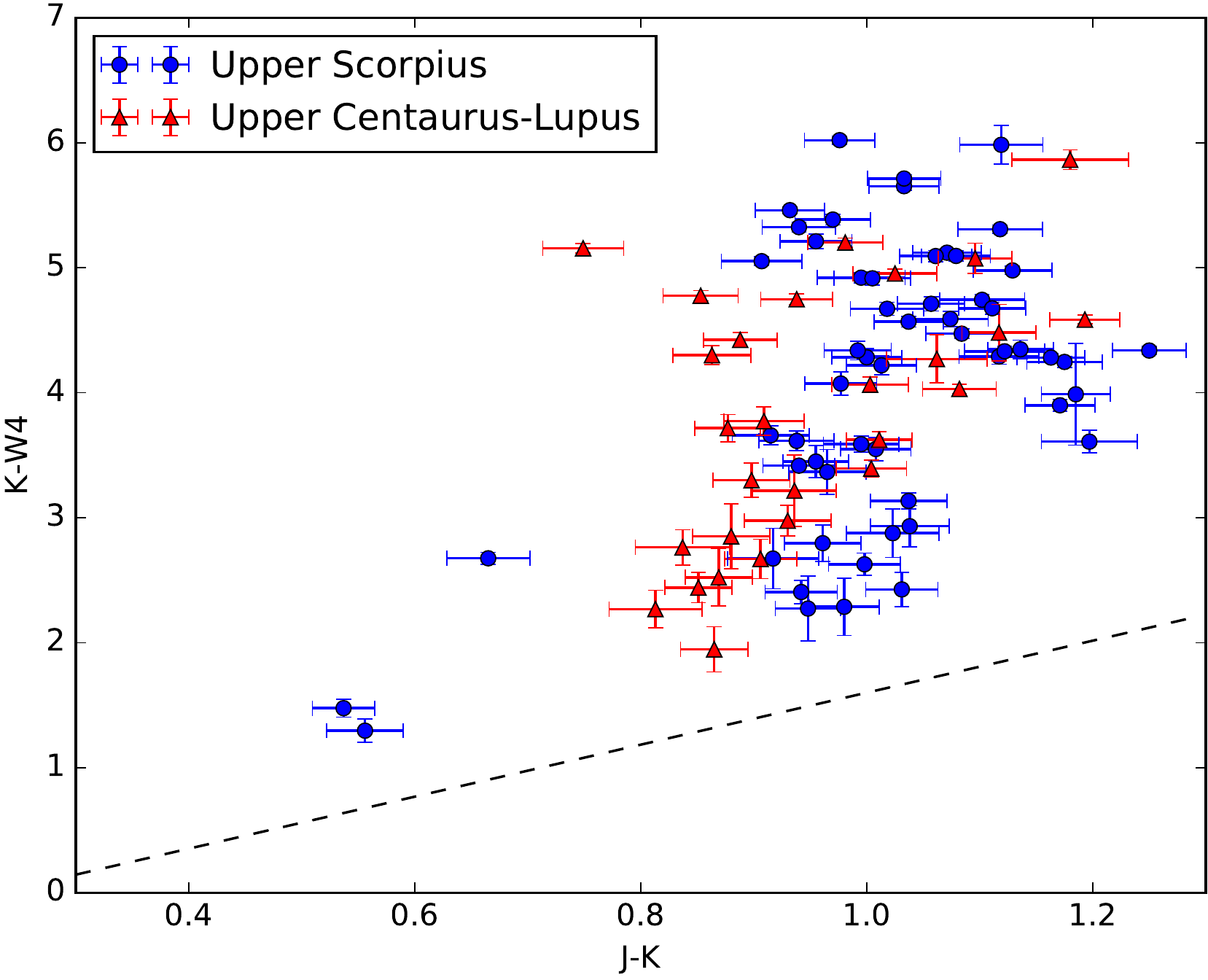}}
	\caption{Colour-colour plot of $K-W4$ against the $J-K$ colours. The dashed lines show the expected $K-W4$ colour based on the photospheric sequence.}
	\label{fig:IRexcess}
\end{figure}

The Wide-field Infra-red Survey Explorer (WISE) telescope carried out an all-sky survey in 4 wavelength bands: W1 ($3.4\,\mu$m), W2 ($4.6\,\mu$m), W3 ($12\,\mu$m) and W4 ($22\,\mu$m) \citep{wright_wide-field_2010}. We cross-matched our objects with WISE IR objects from the AllWISE Source Catalogue. IR excesses in the W2 and W3 bands can indicate the presence of an inner disc and excess in the W4 band would indicate the presence of a colder outer disc. An excess is determined relative to the 2MASS \citep{skrutskie_two_2006} K-band magnitude. The expected $K-W4$ colours based on photospheric sequence of stars without discs is taken to be $(K-W4)=2.08(J-K)-0.48$,  where the $J$ and $K$ magnitudes are taken from the 2MASS catalogue \citep{rizzuto_new_2015}. An object is taken to have an excess if the $K-W4$ colour is at least $3\sigma$ greater than the expected value. These colours are plotted for our targets in \Cref{fig:IRexcess} and their W4 magnitudes are tabulated in \Cref{table:magnitudes1}.

\begin{figure}
	\centerline{\includegraphics[width=1.05\linewidth]{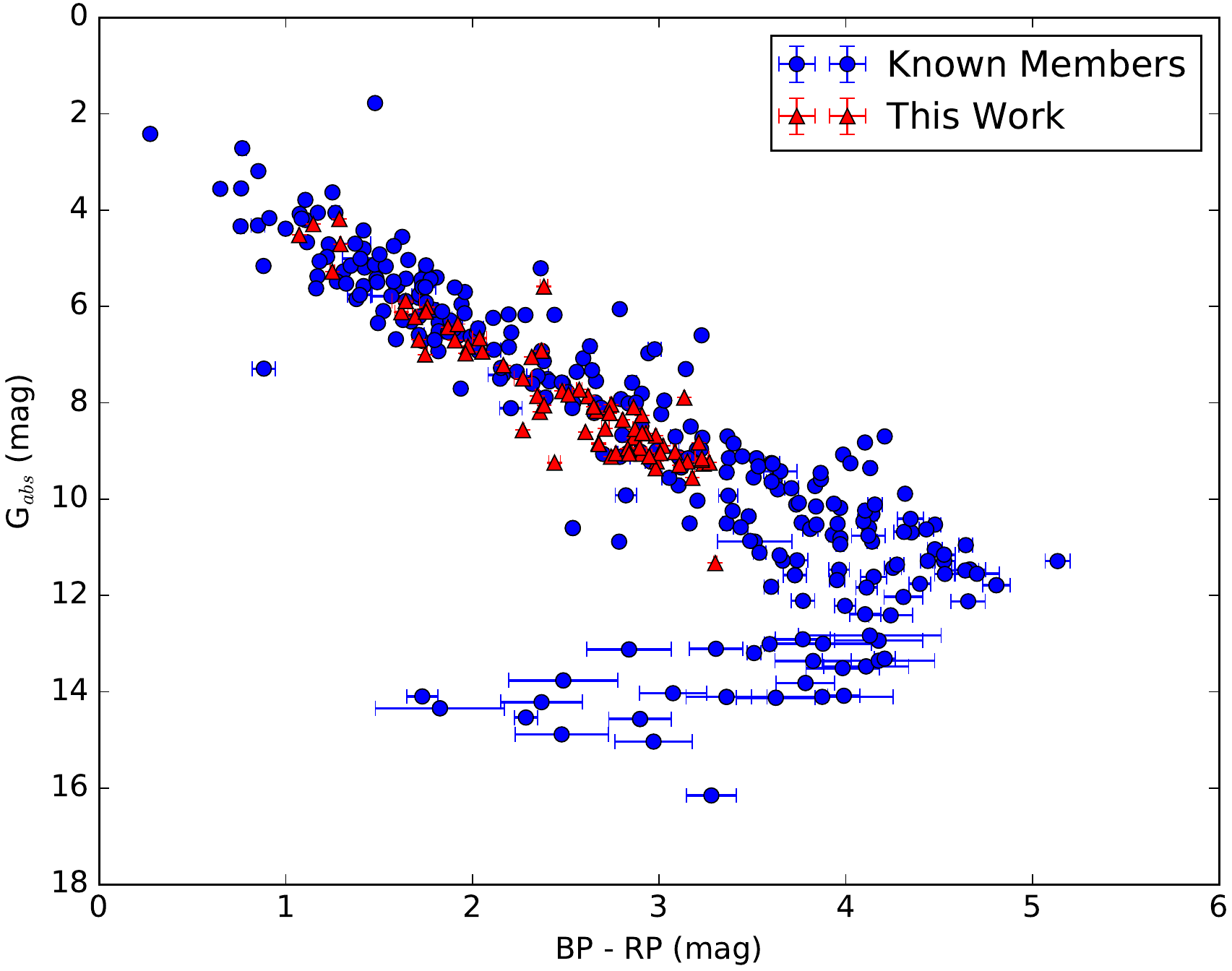}}
	\caption{Colour-magnitude of our targets (red) against previously known members of the Scorpio-Centaurus star forming region (blue). Magnitudes are taken from Gaia (G Mag) and the color is determined from the Gaia Blue band (BP) and Red band (RP) photometry (\citealt{gaia_collaboration_gaia_2016,gaia_collaboration_gaia_2018}). Known members are shown in blue and are taken from \citet{pecaut_star_2016, luhman_disk_2012, aarnio_survey_2008, preibisch_lithium-survey_1998, chen_magellan_2011, martin_spectroscopic_2004, meyer_young_1993, kohler_multiplicity_2000,martin_spectroscopic_1998, bouvier_magnitude-limited_1992,kraus_coevality_2009, lodieu_new_2006, erickson_initial_2011, slesnick_large-area_2006, wilking_optical_2005, oliveira_low-mass_2010, cieza_nature_2010, luhman_low-mass_1999, doppmann_stellar_2003, natta_exploring_2002, elias_infrared_1978, walter_x-ray_1994, ardila_survey_2000,lodieu_multi-fibre_2011, martin_spectroscopic_1998, lodieu_near-infrared_2008, preibisch_exploring_2002, herbig_third_1988, vieira_investigation_2003}. The targets observed in our work is shown in red.}
	\label{fig:CMD}
\end{figure}

\begin{figure}
	\centerline{\includegraphics[width=1.05\linewidth]{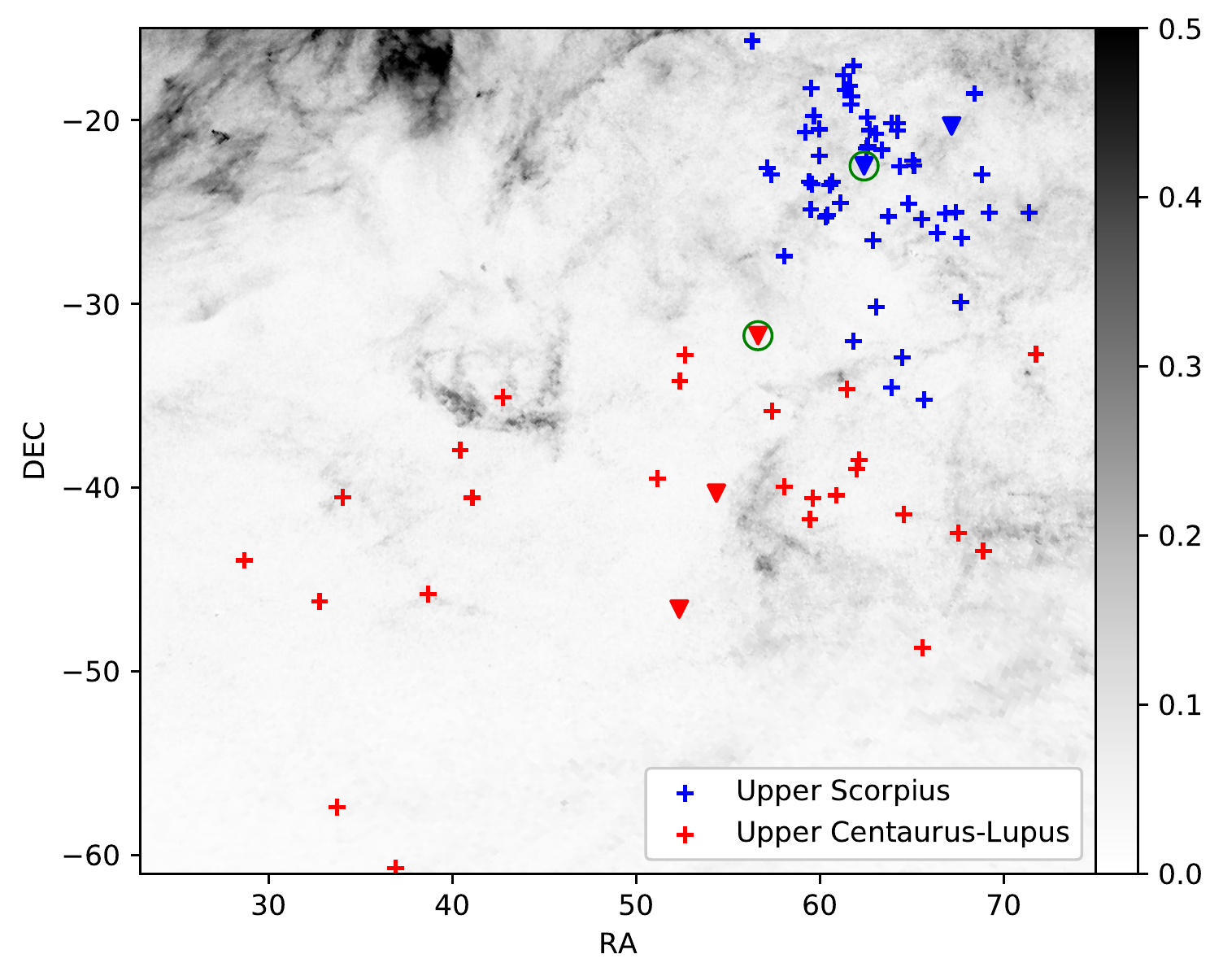}}
	\caption{A map of the candidate disc-bearing stars in Upper Scorpius (US, the \emph{blue} markers) and Upper Centaurus-Lupus (UCL, the \emph{red} markers) that we consider in this study. The background image is an extinction map compiled by \citet{schlafly_map_2014}. The triangle markers indicate objects that we believe to be binary star systems because they have Bayes factor (described by Equation \ref{eqn:bayesfactor} in \Cref{ssec:bayes}) greater than 300. The circled triangles highlight two double-lined spectroscopic binaries found in this work (described in \Cref{ssec:ObjRVs})}
	\label{fig:dust_map}
\end{figure}

IR excess is searched for in our target members of Upper Scorpius and Upper Centaurus-Lupus. These targets were visually inspected on the IRSA WISE image service to check for obscuration by a cloud or contamination by a nearby IR bright object. Some objects were discarded and the final survey used 55 US targets and 28 of UCL targets were found to exhibit IR excesses. The targets are plotted in a colour-magnitude diagram against other known Scorpio-Centaurus members in \Cref{fig:CMD}. The color-magnitude diagram is produced using magnitudes from Gaia (\citealt{gaia_collaboration_gaia_2016,gaia_collaboration_gaia_2018}) and summarised in \Cref{table:magnitudes1}. This shows that our targets are likely to be members of the Sco-Cen star forming region. The positions of our targets are also plotted in \Cref{fig:dust_map} and are summarised in \Cref{table:targets1}. Our targets have large angular separations from the Lupus star-forming clouds. Therefore we believe it is unlikely that we have interlopers from this significantly younger association in our sample. After our disc-bearing targets had been identified they were observed to search for radial velocity (RV) variation which may indicate the presences of a companion.

\subsection{Spectroscopic observations to obtain radial velocities}
\label{ssec:wifes}

\subsubsection{WiFeS observations}
\label{sssec:observations}

To obtain spectroscopic data to determine the radial velocities of our targets, we used the Wide Field Spectrograph (WiFeS) \citep{dopita_wide_2007}. WiFeS is an integral field spectrograph on the Australian National University 2.3m Telescope at Siding Spring Observatory. We use this instrument to search for radial velocity variation over time for our targets. The instrument consists of a blue camera with spectral range $329-558\,\mathrm{nm}$ and red camera with spectral range $529-912\,\mathrm{nm}$.

Data for our targets was collected from June of 2013 to July of 2017. Our observations include Ne-Ar arc exposures every 15-30 minutes to characterise the wavelength scale variation of the instrument over the night due to temperature fluctuations. The WiFeS $R\sim 7000$ grating is used to perform radial velocity measurements. We observe radial velocity standards with well characterised velocities of various spectral types. Radial velocity standards are selected from \cite{nidever_radial_2002} and \cite{soubiran_catalogue_2013} (detailed in Table \ref{table:RVstandards}). Observations of radial velocity standards are bracketed by arc lamp exposures and further correction to the calibrations is applied using oxygen B-band atmospheric absorption lines (static with respect to the observer). The calibration using the oxygen B-band lines is dependent on stellar type and signal-to-noise of each individual target. Data are reduced using the PyWiFeS pipeline\footnote{available at http://pywifes.github.io/pipeline/} \citep{childress_pywifes:_2013}.

\subsubsection{Radial velocity precision based on RV standards}
\label{sssec:rvprecision}

\begin{table}
	\centering
	\begin{tabular}{|c|c|c|c|}
		RV Standard & $v_r$ (km/s) & Sp. Typ. & Approx. Mass (M$_\odot$)\\
		\hline
		HD35974		& 76.683	& G1V$^a$		& 1.40\\
		HD144585	& -14.067	& G2V$^b$		& 1.38\\
		HD145809	& 21.146	& G2V$^c$		& 1.38\\
		HD141885	& -16.160	& G3V$^a$		& 1.37\\
		HD1461		& -10.166	& G3V$^b$		& 1.37\\
		HD189625	& -28.218	& G5V$^d$		& 1.35\\
		HD19034		& -20.344	& G5V$^e$		& 1.35\\
		HD198802	& -3.171	& G5V$^e$		& 1.35\\
		HD153458	& 0.641		& G5V$^e$		& 1.35\\
		HD37213		& 12.464	& G5V$^a$		& 1.35\\
		HD128428	& -42.074	& G6IV$^e$		& 1.34\\
		HD196761	& -41.987	& G7.5IV-V$^b$	& 1.32\\
		HD283		& -43.102	& G9.5V$^c$		& 1.29\\
		HD156826	& -32.634	& K0V$^e$		& 1.26\\
		HD114783	& -12.012	& K1V$^e$		& 1.23\\
		HD32147		& 21.552	& K3+V$^c$		& 1.15\\
		HD2025		& 3.241		& K3V$^c$		& 1.15\\
		HD130992	& -57.160	& K3.5V$^c$		& 1.12\\
		HD170493	& -54.752	& K4V$^e$		& 1.10\\
		HD120467	& -37.806	& K6Va$^b$		& 0.90\\
		GJ173		& -6.768	& M1V$^g$		& 0.55\\
		GJ433		& 17.973	& M1.5$^f$		& 0.50\\
		GJ2066		& 62.205	& M2V$^h$		& 0.45\\
		GJ382		& 7.932		& M2V$^i$		& 0.45\\
		GJ357		& -34.581	& M2.5V$^c$		& 0.40\\
		GJ273		& 18.216	& M3.5V$^i$		& 0.20\\
		GJ729		& -10.499	& M3.5Ve$^j$	& 0.20\\
		\hline
	\end{tabular}
	\caption{Radial velocity standards used as templates to obtain radial velocities of our observed targets. Standards are selected from the \citet{nidever_radial_2002} and \citet{soubiran_catalogue_2013} catalogues. $^a$\citet{houk_michigan_1982}, $^b$\citet{keenan_perkins_1989}, $^c$\citet{gray_contributions_2006}, $^d$\citet{houk_michigan_1988}, $^e$\citet{houk_michigan_1999}, $^f$\citet{henry_solar_2002}, $^g$\citet{stephenson_dwarf_1986}, $^h$\citet{alonso-floriano_carmenes_2015}, $^i$\citet{kirkpatrick_standard_1991}, $^j$\citet{davison_3d_2015}. The approximate masses are derived from the \citet{baraffe_new_2015} pre-main sequence evolution models based on spectral type at age log$_{10}($t$_{age})\sim7.1$.}
	\label{table:RVstandards}
\end{table}

\begin{figure}
	\centerline{\includegraphics[width=1.0\linewidth]{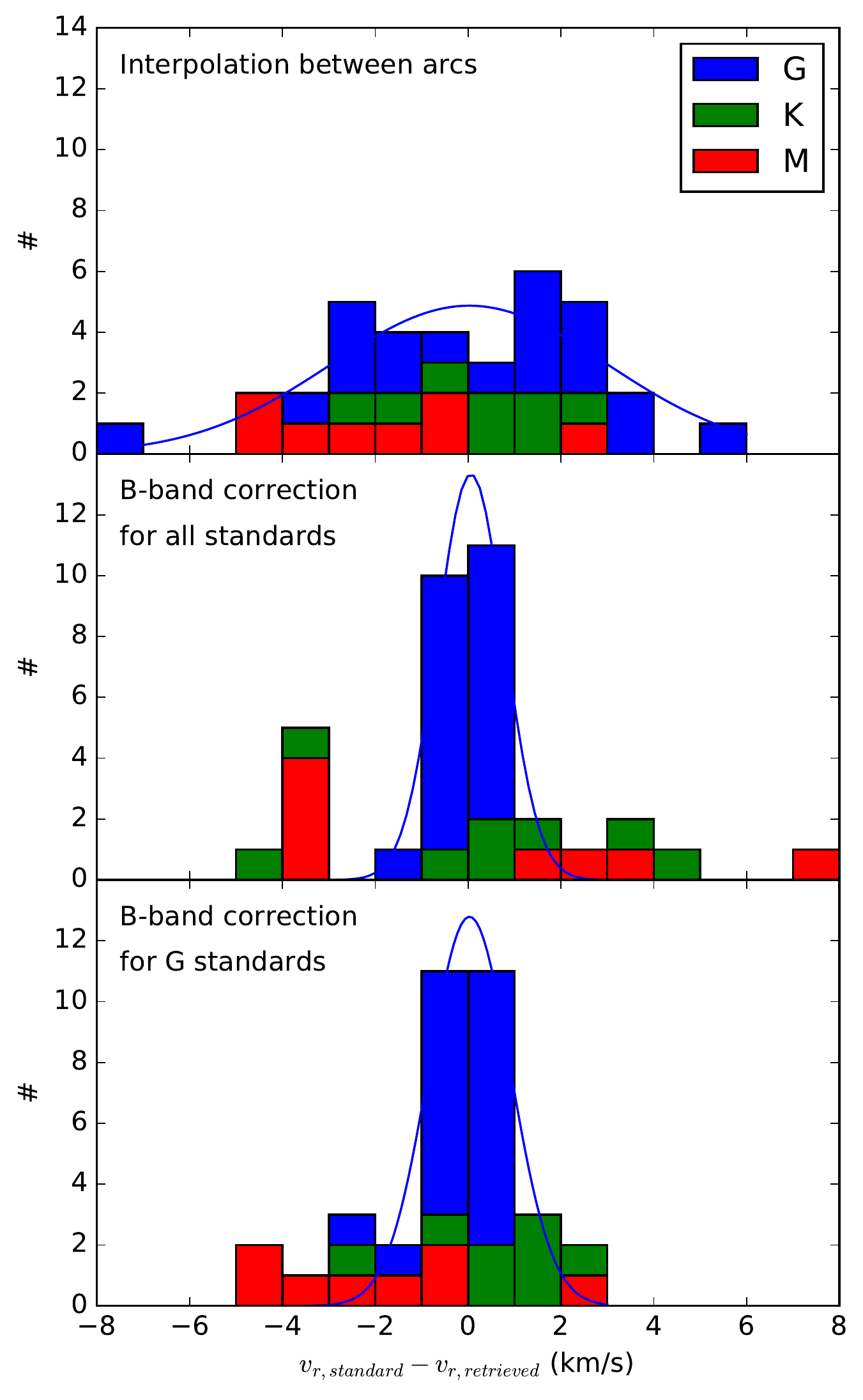}}
	\caption{Precision in retrieving the radial velocity of RV standards. $\Delta v_r$ is the difference between the radial velocity of the RV standard and the retrieved radial velocity using our pipeline. The blue, green, red and magenta histograms show the precision of the G-, K- and M-type RV standards respectively. \emph{Top}: precision when only using the wavelength solutions produced by linear interpolation between arcs (as produced by PyWiFeS). \emph{Middle}: precision when the oxygen B-band absorption lines are also used to further calibrate the wavelength scale. \emph{Bottom}: precision when the oxygen B-band absorption lines are only used for further calibration on the G-type RV standards.}
	\label{fig:RVprecision}
\end{figure}

Radial velocity standards of spectral types varying from G1 through to M3.5 (see Table \ref{table:RVstandards}) were observed with WiFeS to create spectral templates. These templates are cross-correlated with our targets to determine their radial velocity using the post-processing tools developed for WiFeS\footnote{available at http://github.com/PyWiFeS/tools}. Obtained radial velocities are correct to be in heliocentric velocity frame.

The radial velocity standards were observed using the settings described in \Cref{ssec:wifes}. After the observations of the RV standards were reduced, their wavelength scale was shifted to the rest frame. This product was used as the template from which the radial velocity of the targets is obtained.

Tests of retrieving the radial velocity of the RV standards after creating templates of the observed spectra were run. This was done by cross-correlating the reduced spectra of the observed RV standards with created templates, ensuring that the standard was not cross-correlated with itself. We found that the retrieved radial velocities had a relatively large spread from the actual radial velocity of the RV standards with standard deviation of $3.0\,\mathrm{km/s}$ (top panel of \Cref{fig:RVprecision}). We believe this is because the WiFeS instrument may vary on a shorter time-scale than 15-30 minutes. The wavelength scale solution is the linear interpolation between that derived from nearby arcs, as done by the PyWiFeS pipeline, was therefore not accurate enough for the purposes of our work. 

In order to improve our radial velocity precision we use the oxygen B-band atmospheric absorption lines and sky emission. Using the oxygen B-band absorption lines we could determine the offset from the wavelength solution derived from PyWiFeS. The precision of retrieving the radial velocity of the RV standards with the added correction from the B-band is shown in the middle panel of \Cref{fig:RVprecision}. We see that the precision is increased significantly for the G standards, but is decreased for K and M standards. This is not surprising as K and M type stars have emission in the wavelength region of the oxygen B-band absorption, which makes it difficult to determine a velocity. The overall standard deviation of the precision when applying the correction from the B-band absorption reduced to $0.7\,\mathrm{km/s}$.

In addition to using oxygen B-Band absorption lines to improve radial velocity precision, emission from skylines was also tested. This proved to not be useful because the RV standards are very bright, therefore not enough signal-to-noise is obtained to confidently retrieve a radial velocity.

\begin{table}
\centering
\begin{tabular}{|c|c|}
RV standard spectral type & $\sigma$ (km/s)\\
\hline
G	& 0.7\\
K	& 1.4\\
M	& 2.3\\
\hline
\end{tabular}
\caption{Precision of retrieving RV standard radial velocity per spectral type}
\label{table:RVprecision}
\end{table}

\begin{figure}
	\centerline{\includegraphics[width=1.0\linewidth]{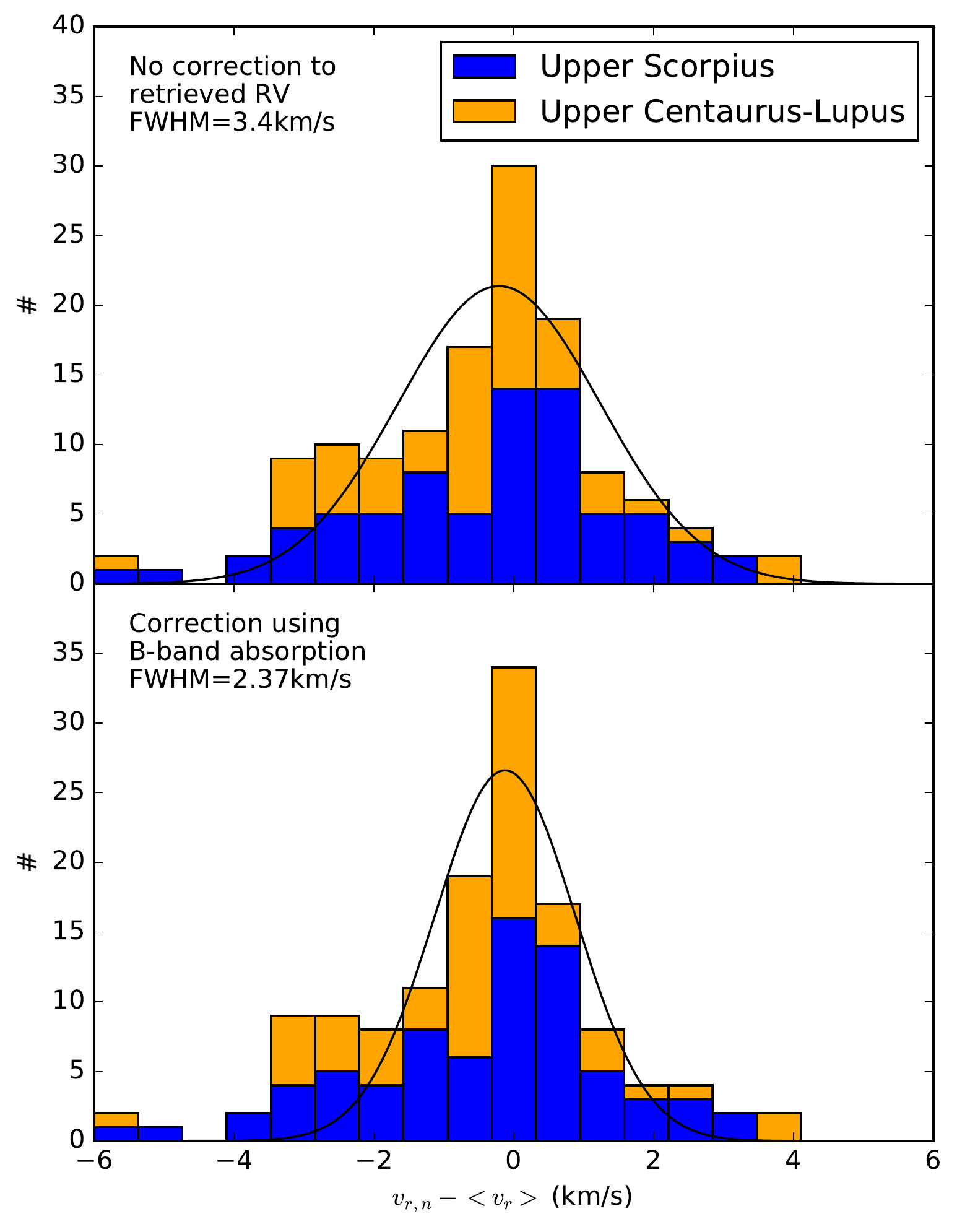}}
	\caption{Radial velocity variation distributions for observed objects calculated as the difference between the radial velocity of the objects last, nth, epoch ($v_{r,n}$) and the objects mean radial velocity over all epochs ($<v_r>$). \emph{Top}: shows the distribution without correction from the absorption lines. The full-width-half-maximum of the fitted gaussian is $3.40$~km/s. \emph{Bottom}: shows the radial velocity variation with the added correction from the absorption lines. The full-width-half-maximum of the fitted gaussian is $2.37$~km/s}
	\label{fig:RVdist}
\end{figure}

Based on the results of this investigation, we decided to only apply calibrations to the G type RV standards by correcting the radial velocity from the shift measured in the B-band absorption. The resulting precision is shown in the bottom panel of \Cref{fig:RVprecision}. The overall standard deviation of the precision using this combination of velocity calibrations is $0.9\,\mathrm{km/s}$, which is a slight increase from applying the B-band absorption correction to all RV standards, but the standard deviation per spectral type is improved. The standard deviation per spectral type is summarised in Table \ref{table:RVprecision}.

\begin{figure*}
	\centering
	\begin{subfigure}[b]{1.0\textwidth}
		\centerline{\includegraphics[width=1.0\linewidth]{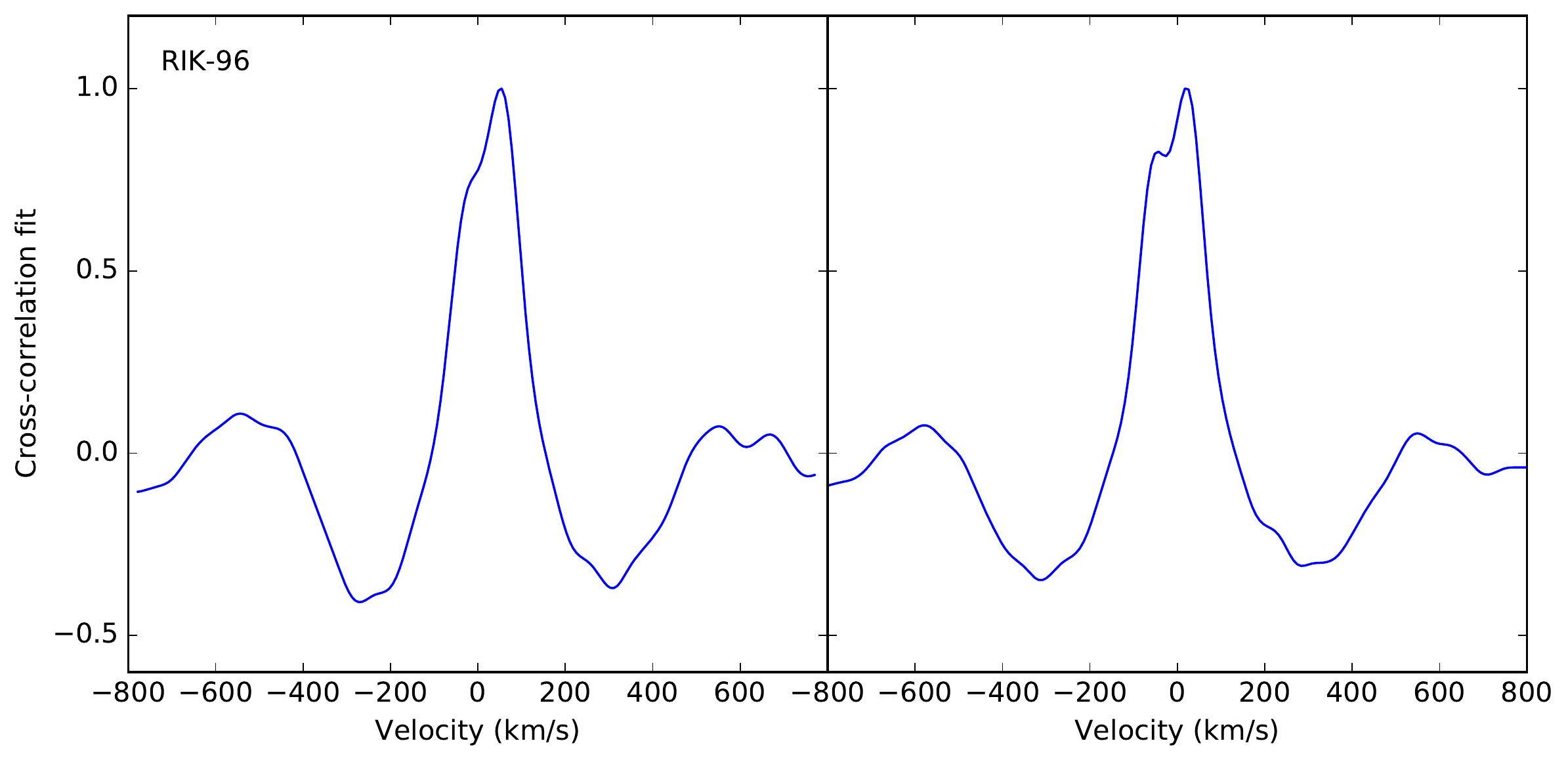}}
	\end{subfigure}
	
	\begin{subfigure}[b]{1.0\textwidth}
		\centerline{\includegraphics[width=1.0\linewidth]{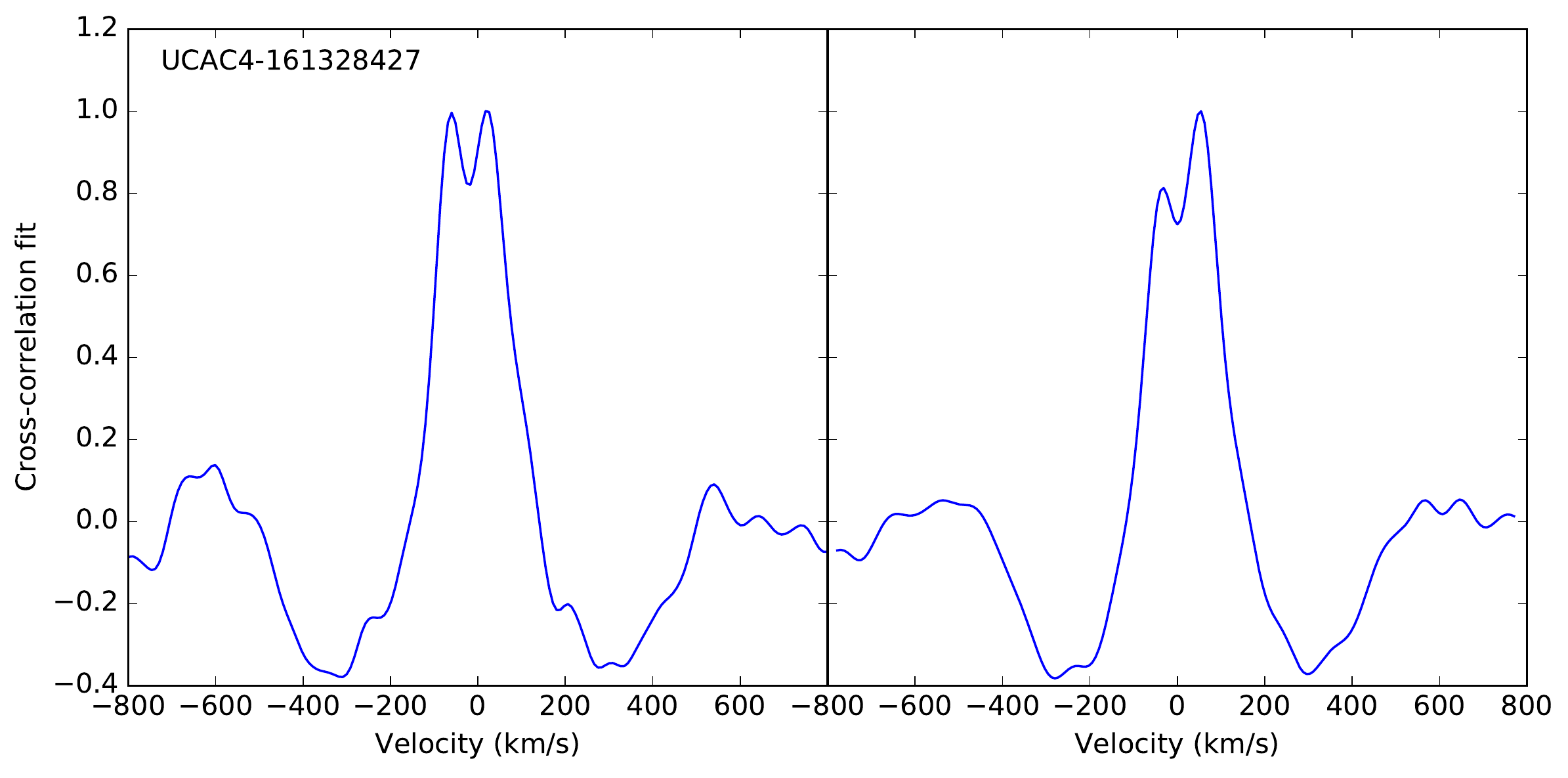}}
	\end{subfigure}
	\caption{Cross-correlations of the two SB2s found. The y-axis has arbitrary units of a measure of fit. The top two panels show cross-correlations for RIK-96 and the bottom two show cross-correlations for UCAC4-161328427 at two different epochs.}
	\label{fig:SB2xcors}
\end{figure*}

\subsubsection{Obtaining radial velocity of targets}
\label{ssec:ObjRVs}

To obtain the radial velocity of our targets, observations are cross-correlated with all the RV standards in an initial step. In order to cross-correlate the target spectra with the template spectra the H$\alpha$ region had to be masked out as most of our targets have strong H$\alpha$ emission (discussed further in \Cref{ssec:Halpha}). Each object is then set a preferred RV standard template to make sure all epochs are cross-correlated with the same template. In the initial step an object will mostly cross-correlate the best with the same template over all epochs or with templates of similar spectral type. An object's preferred template is the template that it correlates the best with over most of the epochs.

After a preferred template is selected, radial velocities are determined again, with all epochs cross-correlating with the same template. Correction using the oxygen B-band absorption is also applied to the obtained radial velocities. The added correction from the absorption lines does show an improvement on the radial velocities precision. When plotting a histogram of radial velocity variation, we expect a large fraction of objects to possess a variation near $0\,\mathrm{km/s}$, which are the objects that are single stars. From our observations, the radial velocity variation is taken to be the difference between the objects last, nth, epoch ($v_{r,n}$) and the objects mean radial velocity over all epochs ($<v_r>$), i.e.

\begin{equation}
\Delta v_r = v_{r,n} - <v_r>.
\label{eqn:deltarv}
\end{equation}

A histogram of radial velocity variation for our targets is shown in \Cref{fig:RVdist}. The full-width-half-maximum (FWHM) of this histogram can be taken as a precision of the radial velocity measurements. We see that the reduces from $3.40\,\mathrm{km/s}$ to $2.37\,\mathrm{km/s}$ when corrections using the oxygen B-band absorption is applied (\Cref{fig:RVdist}).

We identify two double-lined spectroscopic binaries (SB2s). RIK-96 is found in Upper Scorpius and UCAC4-161328427 in Upper Centaurus-Lupus. Cross-correlations at two epochs for each of the SB2s are shown in \Cref{fig:SB2xcors}. Both objects show very strong IR excesses, which likely indicate the presences of a large circumbinary disc. We carried out two dimensional cross-correlations with templates for these objects to estimate the spectral type and flux ratio of the components. We find for RIK-96 the primary and secondary to be of spectral type M1.5 and M2.5 respectively, with a flux ratio of $\sim0.71$. We find for UCAC4-161328427 that both components fit best to M2.5 templates, with a flux ratio of $\sim0.98$. Information about these two objects is summarised in Table \ref{table:SB2s}.

\subsection{H$\alpha$ as a youth indicator}
\label{ssec:Halpha}

\begin{figure}
	\centerline{\includegraphics[width=1.0\linewidth]{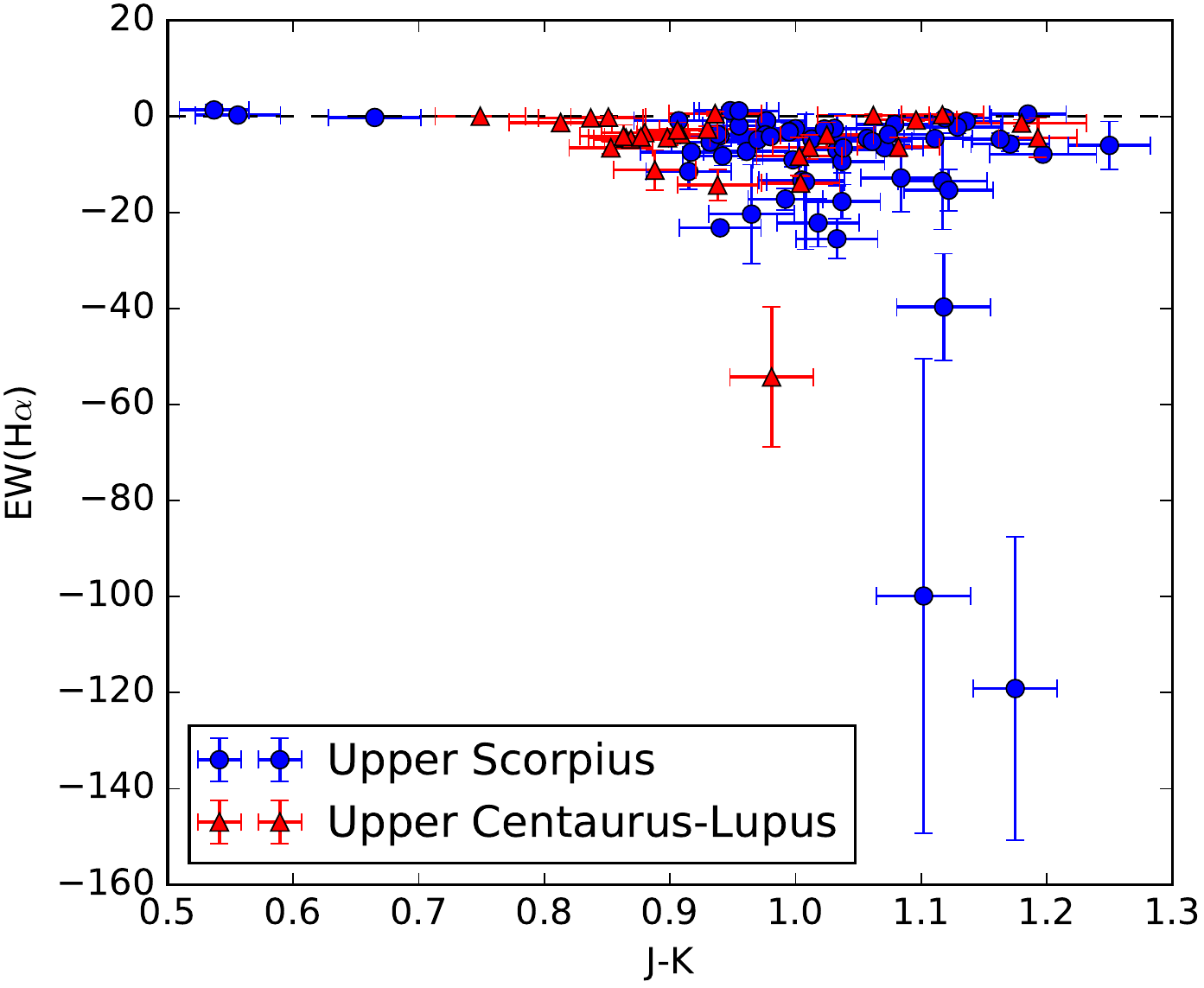}}
	\caption{Mean H$\alpha$ equivalent width for our targets measured over all epochs. The error bars is the standard deviation of the measured equivalent width over all epochs.}
	\label{fig:Halpha}
\end{figure}

As mentioned in \Cref{ssec:ObjRVs} the observed targets showed H$\alpha$ emission that hindered cross-correlations with template spectra. The H$\alpha$ emission is believed to be caused by accretion of disc material onto the stars. In \Cref{fig:Halpha}, we display the measured H$\alpha$ equivalent widths for the targets. The majority of the targets members show some level, of H$\alpha$ emission, with a clear sequence of increasing emission with spectral type. \citet{rizzuto_new_2015} uses the combination of lithium and H$\alpha$ emission as indicator of youth for the identified RIK objects. Of our 55 US and 28 UCL targets, $\sim82\%$ and $\sim70\%$ show H$\alpha$ emission with ($\mathrm{EW(H\alpha)}<-1\AA$). Of the members that do not show or show weak H$\alpha$ emission, most are earlier than M0 spectral type based on the template spectra used for cross-correlation.

\section{Results and Statistical Analysis}
\label{sec:resultsanddiscussion}

Using a hard cut of $5\sigma$ significance on the radial velocity variation for the fraction of binaries in Upper Scorpius and Upper Centaurus-Lupus we get 15$\%$ and 10$\%$ respectively. The fractions produced from this criterion only find the closest binaries, which produce the greatest radial velocity variation. In order to better characterise the binaries that we do not detect we must use Bayesian statistics to explore the parameter space.

\subsection{Simulating Systems}
\label{ssec:simsys}

The radial velocity ($v_r$) of the primary star in a binary system can be described by:

\begin{equation}
v_r = K[\cos(\omega + \Omega) + e \cos\omega] + v_{sys},
\label{eqn:radial_velocity}
\end{equation}

where $K$ is a constant described below, $\omega$ is the longitude of periastron, $\Omega$ is the position angle of the line of nodes, $e$ is the eccentricity of the system and $v_{sys}$ is the system velocity. $K$ is described by:

\begin{equation}
K \equiv \frac{2\pi}{P}\frac{a_1\sin i}{(1-e^2)^{1/2}},
\label{eqn:K}
\end{equation}

where $P$ is the period of the system, $a_1$ is the semi-major axis of the primary star from the barycentre of the system and $i$ is the inclination of the binary. $a_1$ for a system can be calculated for a binary when the period and component masses are known. From the primary mass and mass ratio, the semi-major axis can be determined from Kepler's third law:

\begin{equation}
a = (M_1(1 + q)P^2)^{1/3},
\end{equation}

where $a$ is the semi-major axis of the system, $M_1$ is the mass of the primary star and $q$ is the mass ratio between the secondary and primary components, i.e. $q = M_2/M_1$. From the binary mass function the semi-major axis of the primary component can be determined from:

\begin{equation}
a_1 = a \frac{q}{1 + q}.
\end{equation}

In our Bayesian analysis, we simulate radial velocity curves using $\omega$, $\Omega$, $i$, $e$, $P$, $T_0$, $v_{sys}$, $M_p$ and $q$ sampled from appropriate parameter spaces described in the next section.

\subsection{Bayesian Statistics}
\label{ssec:bayes}

\begin{table}
	\centering
	\begin{tabular}{|c|c|c|}
		& RIK-96 & UCAC4-161328427\\
		\hline
		RA & 16 09 31.66 & 15 46 25.81\\
		DEC& -22 29 22.4 & -31 43 19.3\\
		$J-K$ & 0.995 &0.863\\
		$K-W4$ & 4.922 & 4.305\\
		SpT of Primary & M1.5 & M2.5\\
		SpT of Secondary & M2.5 & M2.5\\
		Flux ratio & 0.71 & 0.98\\
		\hline
		Epoch 1 & & \\
		\hline
		$v_{\mathrm{r,left}}$(km/s) & $-57.5\pm0.3$ & $-63.4\pm0.9$\\
		$v_{\mathrm{r,right}}$(km/s) & $26.5\pm0.2$ & $24.3\pm1.0$\\
		$v_{\mathrm{heliocentre}}$(km/s) & 16.9 & 23.1\\
		\hline
		Epoch 2 & & \\
		\hline
		$v_{\mathrm{r,left}}$(km/s) & $-31.3\pm1.0$ & $-39.5\pm0.7$\\
		$v_{\mathrm{r,right}}$(km/s) & $57.9\pm0.7$ & $56.7\pm0.7$\\
		$v_{\mathrm{heliocentre}}$(km/s) & -10.2 & -5.9\\
		\hline
	\end{tabular}
	\caption{Summary of the SB2s found. The assigned spectral type of the components is based on the best fitting template.}
	\label{table:SB2s}
\end{table}

\begin{table*}
	\centering
	\begin{tabular}{|c|c|c|c|c|}
		Symbol & Parameter & Lower Bound & Upper Bound & Distribution \\
		\hline
		$\omega$ & Longitude of periastron  & $0^{\circ}$ & $360^{\circ}$ & Uniform \\ 
		$\Omega$ & pos. ang. of the line of nodes   & $0^{\circ}$ & $360^{\circ}$ & Uniform\\ 
		$i$ & Inclination  & $0^{\circ}$ & $90^{\circ}$ & Sinusoidal\\
		$e$ & Eccentricity  & $0$ & $0.95$ & Uniform\\ 
		$P$ & Period & $1\,\mathrm{day}$ & $7300\,\mathrm{days}$ & Log normal\\
		$T_0$ & Time of Periastron Passage & $0\,\mathrm{days}$ & $P$ & Uniform\\
		$v_{sys}$ & system velocity & $<v_r> - 4\,\mathrm{km/s}$&$<v_r> + 4\,\mathrm{km/s}$ & Gaussian\\
		\hline
		$M_p$ &Mass of Primary & $M_{template} - 0.1M_{template}$ & $M_{template} + 0.1M_{template}$ & Uniform\\ 
		$q$ & mass ratio & $0.05$ & $1.0$ & Uniform\\
	\end{tabular}
	\caption{Priors for Bayesian analysis.}
	\label{table:priors}
\end{table*}

Some of these priors are independent from the physical characteristics of the system, such as $\omega$, $\Omega$ and $i$, therefore the parameter space from which these quantities are sampled is the same for all objects. The parameter spaces that the period, time of periastron passage and eccentricity are sampled from also remain the same for all objects because these quantities directly influence the shape of the radial velocity curve. The parameter space sampled for the system velocity is dependent on the observed radial velocities, therefore this space is redefined for each object, as described in \Cref{ssec:priors}. 

Initially all random orbits are given an amplitude of $1\,\mathrm{km/s}$, but this is scaled when considering the possible mass of the primary star and the mass ratios. For each object $\omega$, $\Omega$, $i$, $P$, $T_0$, $e$ and $v_{sys}$ are sampled $10^5$ times to produce radial velocity curves with normalised amplitudes. Then for each of these orbits $M_p$ and $q$ are sampled $10^2$ times to scale these normalised radial velocity curves. The parameter space that the primary mass is sampled is dependent on the template that the object cross-correlates with. The parameter space that the mass ratio is sampled from is the same for all objects.

Once a radial velocity curve is simulated, a likelihood of fit is calculated for each simulated orbit given by:

\begin{equation}
L \propto e^{-\frac{(v_{r,obs} - v_{r,sim})^2}{2\sigma_{obs}^2}},
\end{equation}

where $v_{r,obs}$ is the observed radial velocity, $v_{r,sim}$ is the simulated radial velocity, and $\sigma_{obs}$ is the error on the measured radial velocity. The error of the measured radial velocity is the error derived from the cross-correlation, the error from the precision of the template used and a standard astrophysical error added in quadrature. The standard astrophysical error accounts for variations caused by stellar activity or rotation and is set to $1\,\mathrm{km/s}$. Our results are shown not to vary significantly for an astrophysical error over the range of $0.5-2.0\,\mathrm{km/s}$. The error on the precision of the RV standard template is taken from Table \ref{table:RVprecision}. 

Likelihoods are also calculated for fitting a single star i.e. a constant radial velocity. The velocity for a single star fit is taken to be the system velocity ($v_{sys}$) that is sampled for the simulated binary star orbit.

After all the radial velocity curves for the simulated orbits have likelihoods calculated, we calculate the mean likelihood of the data given our two models, $P(D|S)$ and $P(D|B)$. From these, the Bayes factor is calculated, which is given by:

\begin{equation}
\Upsilon = \frac{P(D|B)}{P(D|S)},
\label{eqn:bayesfactor}
\end{equation}

From this we see that the Bayes factor is simply the ratio of the mean likelihoods of being a binary star to a being a single star. A binary model is preferred if $\Upsilon \gg 1$ and a single star model is preferred if $\Upsilon \ll 1$. Where $\Upsilon \approx 1$, either model is equally probable. We take objects with Bayes factors $> 300$ to be binary stars as this is near the $3\sigma$ confidence level assuming a Gaussian probability distribution. Using this threshold we find five objects to be binary stars, including the two double-lined binaries shown in \Cref{fig:SB2xcors} and listed in \Cref{table:SB2s}. The Bayes factors for every target can be found in \Cref{table:targets1}

Once the Bayes factor is calculated for each target we calculate the binary fraction. The likelihood of being a binary or multiple star system, $P(B)$, is complimentary to the likelihood of being a single star, $P(S)$. Therefore, if we take the likelihood of being a binary to be $\gamma$, then $P(B) = \gamma$ and $P(S) = 1 - \gamma$. This is described statistically like a \emph{Bernoulli trial}. The Jeffreys' prior to this case states that the probability of $\gamma$ can be described by:

\begin{equation}
P(\gamma) \propto \frac{1}{\gamma(1-\gamma)}.
\label{eqn:jeffreysprior}
\end{equation}

We are trying to find what the probability density function of $\gamma$ is for Upper Scorpius and Upper Centaurus-Lupus. Following Bayes' Theorem, the probability of $\gamma$ given our data is:

\begin{equation}
P(\gamma|D) = \frac{P(\gamma)}{P(D)}P(D|\gamma).
\label{eqn:bayestheorem}
\end{equation}

We can expand out the probability of the data given $\gamma$, $P(D|\gamma)$, into constituents of the probability of being a single or binary star given $\gamma$, $P(S|\gamma)$ and $P(B|\gamma)$ respectively, giving:

\begin{equation}
P(\gamma|D) = \frac{P(\gamma)}{P(D)}[P(D|B)P(B|\gamma) + P(D|S)P(S|\gamma)].
\end{equation}

As $\gamma$ is defined as the multiplicity, $P(B|\gamma) \equiv \gamma$ and $P(S|\gamma) \equiv 1-\gamma$ as described previously. This gives us:

\begin{equation}
P(\gamma|D) = \frac{P(\gamma)}{P(D)}[P(D|B)\gamma + P(D|S)(1-\gamma)].
\label{eqn:bayesintermediate}
\end{equation}

We can express Equation \ref{eqn:bayesintermediate} in terms of our Bayes' factor, which we calculate from our observations. The probably of $\gamma$ given the data is now:

\begin{equation}
P(\gamma|D) = \frac{P(\gamma)}{P(D)}P(D|S)[\Upsilon\gamma + (1-\gamma)].
\label{eqn:gammaprobability}
\end{equation}

Or to simplify:

\begin{equation}
P(\gamma|D) \propto P(\gamma)[\Upsilon\gamma + (1-\gamma)].
\label{eqn:gammaprobabilitysingle}
\end{equation}

Where $P(\gamma)$ is our prior on the probability distribution of the binary fraction. Our initial guess is that the probability distributions of $\gamma$ is uniform, meaning that every possible binary fraction is equally likely. This prior is updated with the Bayes' factor of each object. For the first object, the probability of $\gamma$ given the data is:

\begin{equation}
P(\gamma|D)_1 \propto P(\gamma)_0[\Upsilon_1\gamma + (1-\gamma)].
\label{eqn:gammaprobabilityfirst}
\end{equation}

Where $P(\gamma)_0$ is our initial guess of a uniform distribution. This is saying that we believe this object has an equal probability of being a single or binary star and it is updated with the Bayes factor of the first object. This updated prior becomes the prior for the next object. Expanding this to $n objects$ we get:

\begin{equation}
P(\gamma|D) \propto P(\gamma) {\displaystyle \prod_{i=1}^{n}} [\Upsilon_i\gamma + (1-\gamma)].
\label{eqn:gammaprobabilityn}
\end{equation}

\subsection{Bayesian priors}
\label{ssec:priors}

The priors for our Bayesian analysis are summarised in Table \ref{table:priors}. $\omega, \Omega$ and $i$ are all independent of physical characteristics of the binary star system. These quantities account for projection effects on the observed radial velocity and may have any orientation, i.e. are isotropic. We sample inclination $i$ from a distribution proportional to sin($i$), and sample the position angle of the line of nodes, $\Omega$, and longitude of periastron, $\omega$, from 0$^\circ$ to 360$^\circ$ in order to uniformly sample the three-dimensional orientation of orbits.

The eccentricity of an object may be $e=0$ for a circular orbit, up to $e=1$ for a parabolic orbit. We sample the eccentricity from a uniform distribution as \cite{raghavan_survey_2010} find that eccentricity follows a mostly uniform distribution up to $e=0.6$. The lower bound for our eccentricity parameter space is $e=0$ and the upper bound is $e=0.95$. 

The largest baseline our observations have is $\sim$4$\,\mathrm{yrs}$. With this baseline we take the largest binary orbital period that we should be able to detect to be $7300\,\mathrm{days}$ or $\sim$20$\,\mathrm{yrs}$. We take the shortest possible period to be $1\,\mathrm{day}$. The periods are sampled from a log-normal distribution with a mean of $\mu_P = 5.03$ and standard deviation $\sigma_P = 2.28$ as found by \cite{raghavan_survey_2010}. The time of periastron passage, $T_0$, is essentially the phase of the radial velocity curve and is dependent on the period. All values of $T_0$ are equally likely, so this value is sampled from a uniform distribution from $0$~days to $P$ in days.

We aim to marginalise our probability distributions over all possible values of the centre of mass velocity, because the dispersion of velocities within our association is larger than our typical velocity precision. For computational speed, we only sample values of $v_{sys}$ from a normal distribution centred on the mean radial velocity over all epochs of an object and with a standard deviation of 4$\,\mathrm{km/s}$ (larger than our typical uncertainties). Sampling from a wider distribution would lower both the single and binary star probabilities roughly equally and would not noticeably affect our results.

It is reasonable to assume that the best fitting spectral template corresponds to the brighter component of a binary star system. From this assumption we sample the mass of the primary, $M_p$, based on the template that was used to derive radial velocities. The typical mass of a main sequence star for the spectral type of the template is used as a guide for $M_p$. The typical spectral type mass for each template can be found in Table \ref{table:RVstandards}. $M_p$ is sampled from a uniform distribution with lower and upper bounds being $\pm 10\%$ of the typical mass of the template. The mass ratio, $q$, is sampled from a uniform distribution as this is what observations have shown \citep{raghavan_survey_2010}. For our work we set a minimum mass ratio of $q= 0.05$.

\begin{figure}
	\centerline{\includegraphics[width=1.0\linewidth]{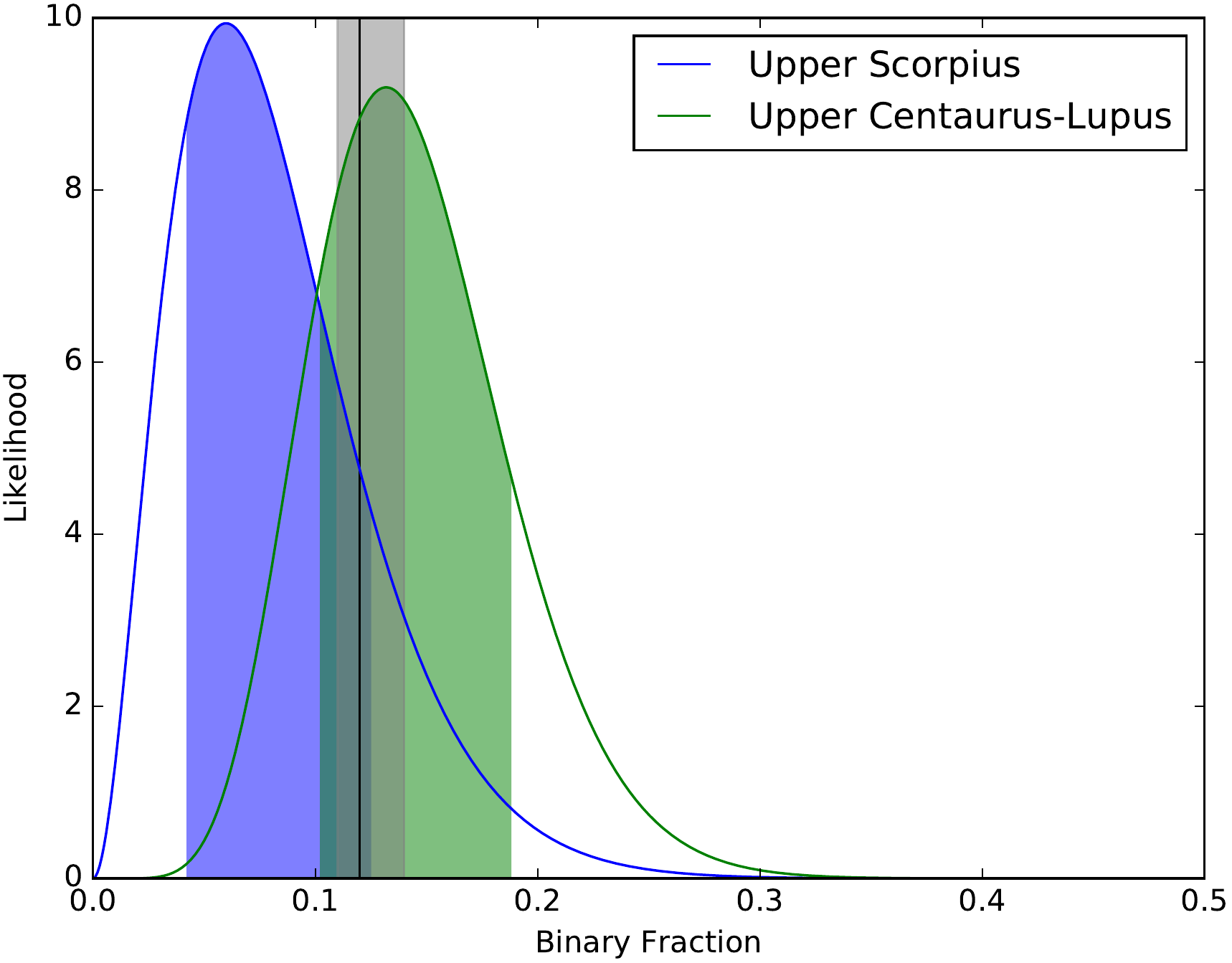}}
	\caption{The normalised probability distribution of multiplicity for our disc-bearing stars in Upper Scorpius and Upper Centaurus-Lupus. The y-axis is in arbitrary units. The vertical line and shaded region highlights our expected binary fraction of $\sim0.12^{0.02}_{0.01}$ based on the multiplicity of field stars irrespective of whether they have discs or not. The mean likelihood fractions of US and UCL are $0.06^{0.07}_{0.02}$ and $0.13^{0.06}_{0.03}$ respectively.}
	\label{fig:binaryfraction}
\end{figure}


When carrying out this calculation for our objects in Upper Scorpius and Upper Centaurus-Lupus, we find that the maximum likelihood binary fractions to be $0.06^{0.07}_{0.02}$ and $0.13^{0.06}_{0.03}$ respectively. The mean likelihoods for Upper Scorpius and Upper Centaurus-Lupus are $0.08$ and $0.14$ respectively. The normalised posterior density functions are shown in \Cref{fig:binaryfraction}. The error bars are the 16th and 84th percentile, and are highlighted in \Cref{fig:binaryfraction}.

\section{Discussion}
\label{sec:discussion}

\subsection{Binary fraction and implication on disc lifetime}
\label{ssec:disc_lifetime}

From our work we find that approximately one fifth of disc-bearing stars of spectral types F and later in Upper Scorpius (US) and Upper Centaurus-Lupus (UCL) are in binary star systems with periods less than 20 years. 

\cite{raghavan_survey_2010} find the multiplicity of $\sim1.25-1.02$~M$_\odot$ stars to be $54\%\pm4\%$ and $\sim1.02-0.78$~M$_\odot$ stars to be $41\%\pm3\%$. \citet{shan_multiplicity_2017} find a raw multiplicity of $\sim0.9-0.07$~M$_\odot$ dwarfs to be $35\%\pm3\%$. From these multiplicities found in the literature and the masses of our targets (based on the template fitted to the target, see \Cref{ssec:ObjRVs}) we calculate an expectation value for the binary fraction based on the field population. Based on the templates fit to our targets we have 6 US and 4 UCL objects $\ge1.02$~M$_\odot$, 7 US and 8 UCL objects between $1.02-0.78$~M$_\odot$ and 42 US and 16 UCL objects below $\sim0.78$~M$_\odot$. Under the assumption that binary stars of all spectral types follow the same log-normal orbital period distribution as described by \cite{raghavan_survey_2010}, we are able to detect up to $31\%$ of all binary systems (because $31\%$ of the log-normal period distribution is below a period of 20 years). Based on the binary fractions and the fraction of detectable binary star systems, we expect $\sim$1 US and UCL target $\ge1.02$~M$_\odot$, $\sim$1 US and UCL between $1.02-0.78$~M$_\odot$ targets, and $5\pm0.5$ of our US and $\sim2$ of our UCL objects below $0.78$~M$_\odot$ to be binaries. This gives a total expectation value of $7\pm0.5$ out of the 55 US targets and $3.5\pm0.5$ out of the 28 UCL targets to be binaries, or an expected fraction of $\sim0.12^{0.02}_{0.01}$ for both regions if they followed the same binary statistics as the field stars in \cite{raghavan_survey_2010}. This is consistent with our fractions obtained from the Bayesian analysis.

Our results show that the binary fraction of our targets showing IR excess is tentatively consistent with the binary fraction of the general stellar population implying that the overall the lifetime of discs in binary star systems is comparable to single stars. Our work does not discriminate between circumbinary discs and circumstellar discs around individual components in a binary star system. There may be variations in disc lifetime depending on the configuration of the system. \cite{damjanov_comprehensive_2007} and \cite{harris_resolved_2012} find that binaries of separation $\lesssim$30$\,\mathrm{AU}$ have significantly smaller IR excess compared to other binary stars with discs, which may indicate that a close companion leads to faster disc dispersal. This is in contrast to many massive, bright circumbinary discs (\citealt{harris_resolved_2012, cox_protoplanetary_2017}) which could indicate the opposite. A binary star has more angular momentum compared to a single star of the same total mass, so one would expect a binary star system to host a larger disc than it's single star counterpart. The primary of a binary star system would also accrete less rapidly than a single star of the combined mass of binary components, which would lead to a slower photo-evaporation process, leading towards a longer lifetime. \citet{daemgen_frequency_2016} find that the fraction of accretion discs around binary stars is a factor of $\sim2$ lower than accretion discs around single stars in the Chamaeleon I region. This may be because circumbinary discs have the gas evacuated from the inner region by a close companion, leading to lower accretion rates. This could contribute to circumbinary discs having a longer typical lifetime and explain some very old circumbinary discs. These various factors could average out over the long term leading to similar disc lifetimes in single star and binary star systems.




\subsection{Caveats}
\label{ssec:caveats}

In this work we use Bayesian statistics to account for the limited set of observations on this study. As such, we must discuss some of the caveats  associated with our conclusions.

Only the two double-lined spectroscopic binaries presented in \Cref{fig:SB2xcors} can be conclusively identified as binaries. For the remainder of our objects, binarity is determined from a combination of a detected radial velocity variation and Monte-Carlo simulations. This is because no single object has had a sufficient number of radial velocity observations to resolve a full orbit, with only 2-7 epochs, with a median of 4 epochs per target.

Additionally, given we have only two conclusively identified binary systems, our work gives a raw binary fraction which is consistent with no multiplicity in the sample. Therefore, the suggestion that discs in binary star systems are heavily hindered leading to the quick dispersal of discs in binaries of periods up to 20 years is also valid based on our data. 

The $22\mu$m emission detected by the WISE W4 band traces disc emission out to a few AU, whereas millimetre surveys from \citet{harris_resolved_2012} and \citet{cox_protoplanetary_2017} trace regions that are further out. This makes comparison with previous work difficult as the infrared emission measured in this work probes a different region of the IR spectrum. Our work looks at the survival time of the inner region of discs, which may be truncated in circumbinary discs. But these same circumbinary discs would, nevertheless, exhibit millimetre emission.

Our target sample is also biased against massive stars (i.e.
greater than $\sim1.4$~M$_\odot$). This is mainly due to a lack of absorption lines needed for precise cross-correlation with a template. This may be problematic in the context of determining multiplicity as many surveys \citep{raghavan_survey_2010} show that multiplicity increases with primary star mass. But the overall multiplicity of disc-bearing objects would be more sensitive to the initial mass function used and the populations of lower-mass stars. Therefore we believe that excluding massive stars is unlikely to change our results significantly.

\section{Summary and Conclusion}
\label{sec:conclusion}

We determine the multiplicity of disc-bearing stars in the $\sim$11$\,\mathrm{Myr}$ Upper Scorpius (US) and $\sim$17$\,\mathrm{Myr}$ Upper Centaurus-Lupus (UCL) for periods up to 20 years. Our sample consists of 55 US members and 38 UCL members that are shown to have an IR excess, indicating the presence of a disc. Targets are observed with the Wide Field Spectrograph (WiFeS) on the Australian National University 2.3m telescope to search for radial velocity variation.

Using Bayesian statistics we determine the binary fraction of our disc-bearing stars in Upper Scorpius and Upper Centaurus-Lupus to be $0.06^{0.07}_{0.02}$ and $0.13^{0.06}_{0.03}$ respectively. When compared to the expected binary fraction of $\sim0.12^{0.02}_{0.01}$ based on our observational limits and the multiplicity of field stars, these results are consistent with disc lifetime around binary stars being similar to single stars. This implies that planet formation is equally likely around binary stars as around single stars.

\section*{Acknowledgements}

The authors would like to thank the anonymous referee for their insightful comments and suggestions. R.K.~would like to thank the Australian Government and the financial support provided by the Australian Postgraduate Award. R.K.~also acknowledges the time awarded on the ANU 2.3m telescope on proposals 1150125, 2150198, 3150150, 1160210, 2160158, 2160114, 3160154 and 2170107. We acknowledge the traditional owners of the land on which the ANU 2.3m telescope stands, the Gamilaraay people, and pay our respects to elders, past and present. This research has made use of the AllWISE Source Catalog created from NASA/IPAC Infra-red Science Archive, which is operated by the Jet Propulsion Laboratory, California Institute of Technology, under contract with the National Aeronautics and Space Administration. M.I.~thanks the Australian Research Council for grant FT130100235. C.F.~gratefully acknowledges funding provided by the Australian Research Council's Discovery Projects (grants~DP150104329 and~DP170100603) and by the Australia-Germany Joint Research Co-operation Scheme (UA-DAAD). Bayesian calculations were carried out on high performance computing resources provided by the Leibniz Rechenzentrum and the Gauss Centre for Supercomputing (grants~pr32lo, pr48pi and GCS Large-scale project~10391), the Partnership for Advanced Computing in Europe (PRACE grant pr89mu), the Australian National Computational Infrastructure (grant~ek9), and the Pawsey Supercomputing Centre with funding from the Australian Government and the Government of Western Australia, in the framework of the National Computational Merit Allocation Scheme and the ANU Allocation Scheme.

\bibliographystyle{mn2e}
\bibliography{Bibliography}

\appendix
\section{Targets}
\label{sec:appendix}

Listed in \Cref{table:magnitudes1} is the $J$ and $K$ magnitudes for our targets taken from the 2MASS survey \citep{skrutskie_two_2006}. The W4 magnitude is take from the AllWISE source catalogue \citep{wright_wide-field_2010}. We use Gaia (\citealt{gaia_collaboration_gaia_2016,gaia_collaboration_gaia_2018}) to obtain Gaia magnitudes (G Mag) as well as Blue band (BP) and Red band (RP) photometry.

Listed in \Cref{table:targets1} is the full list of targets observed in this survey.

The probability of membership, P(M), is calculated based on the calculations of \citet{rizzuto_multidimensional_2011} and \citet{rizzuto_new_2015}  which uses kinematic and spatial information to assign a probability (see \Cref{ssec:membership}).


$\Delta v_r$ is the observed radial velocity variation over all epochs. The radial velocity variation is taken to be the difference between the epoch with the highest radial velocity and the epoch with the lowest radial velocity, (see Equation \ref{eqn:deltarv}).

$\Upsilon$ is the Bayes factor calculated as described in \Cref{ssec:bayes}. A binary star model is preferred if $\Upsilon \gg 1$ and a single star model is preferred if $\Upsilon \ll 1$. Where $\Upsilon \approx 1$, either model is equally probable. We take objects with Bayes factors $> 300$ to be binary stars as this is near the $3\sigma$ confidence level assuming a Gaussian probability distribution.

\begin{table*}
	\centering
	\caption{Magnitudes of our targets. J and K magnitudes are taken from the 2MASS survey. W4 magnitude is taken from the AllWISE source catalogue. The G, BP and RP magnitudes are taken from the Gaia survey.}
	\begin{tabular}{|c|c|c|c|c|c|c|c|c|}
		Object & RA & DEC & $J$ & $K$ & $W4$ & $G$ & $BP$ & $RP$\\
		\hline
		UCAC4-827387234&13 54 35.61&-43 56 48.1&11.638&10.708&7.732&14.46715&15.951675&13.276466\\
		UCAC4-467124593&14 10 59.69&-46 11 33.8&11.196&10.343&5.567&14.351444&16.073818&13.087802\\
		UCAC4-947158099&14 14 52.40&-57 22 54.0&13.262&12.2&7.93&15.941499&17.134708&14.865117\\
		UCAC4-382098779&14 16 03.10&-40 30 34.2&11.014&10.149&8.202&13.938769&15.460843&12.718367\\
		UCAC4-945204860&14 27 30.11&-60 42 18.0&12.588&11.492&6.418& & & \\
		UCAC4-442012351&14 34 36.62&-45 45 59.9&11.725&10.845&7.995&14.812784&16.228548&13.460551\\
		UCAC4-371257082&14 41 37.66&-37 56 34.4&11.746&10.837&7.065&14.544952&15.975403&13.370145\\
		UCAC4-371337294&14 44 15.62&-40 32 27.3&11.654&10.766&6.344& & & \\
		UCAC4-21594717&14 50 55.74&-35 04 22.9&10.892&10.055&7.293&12.948328&13.802399&12.056115\\
		UCAC4-365499547&15 24 35.62&-39 30 43.3&11.56&10.691&8.169&14.581901&16.145092&13.340626\\
		UCAC4-445278523&15 29 17.48&-46 36 28.7&12.51&11.574&8.358&13.758428&14.996365&12.6358185\\
		UCAC4-23251943&15 29 23.88&-34 11 54.5&11.535&10.637&7.338& & & \\
		UCAC4-23288203&15 30 33.57&-32 46 19.1&11.08&10.174&7.504& & & \\
		UCAC4-447414452&15 37 22.69&-40 17 59.6&8.796&7.771&2.818&10.21763&10.68769&9.616082\\
		UCAC4-333669217&15 45 09.50&-15 40 29.0&11.808&10.795&6.576&15.126357&17.03052&13.792687\\
		UCAC4-161328427&15 46 25.81&-31 43 19.3&11.468&10.605&6.305&14.510724&16.173733&13.265781\\
		UCAC4-30196323&15 48 24.44&-22 35 49.7&11.868&10.732&6.385&15.303416&17.155664&13.978409\\
		UCAC4-30231541&15 49 19.76&-22 57 29.7&10.534&9.536&6.909&13.884282&15.718709&12.584246\\
		UCAC4-1297314504&15 49 30.74&-35 49 51.4&10.96&10.022&5.274&13.875682&15.305055&12.686407\\
		UCAC4-408863727&15 52 08.84&-27 23 45.8&11.996&11.019&6.946&15.29399&16.951279&14.044705\\
		UCAC4-1019544336&15 52 13.44&-39 56 08.2&11.705&10.828&7.112&14.595463&16.07941&13.4024315\\
		RIK-23&15 56 42.45&-20 39 34.2&11.313&10.28&4.628&14.495734&16.086412&13.223093\\
		RIK-30&15 57 34.31&-23 21 12.3&9.932&8.992&5.576&12.7145195&13.970277&11.600752\\
		UCAC4-519496506&15 57 43.61&-41 43 37.8&9.825&8.743&4.713&12.037338&12.895082&11.136776\\
		UCAC4-426579903&15 57 54.45&-24 50 42.4&9.51&8.578&3.118&11.776407&12.5178175&10.898948\\
		UCAC4-404438620&15 57 58.92&-18 14 59.6&9.991&8.92&3.8&12.300475&13.221161&11.352783\\
		RIK-34&15 58 12.70&-23 28 36.4&8.574&8.018&6.722&10.07295&10.586112&9.43927\\
		UCAC4-519508031&15 58 20.40&-40 33 05.8&10.563&9.582&4.381&13.720713&14.981765&12.501845\\
		RIK-38&15 58 36.20&-19 46 13.5&11.697&10.721&4.701&14.877639&16.626127&13.604954\\
		RIK-43&15 59 44.27&-20 29 23.4&11.522&10.405&6.116&14.968601&16.911917&13.642211\\
		UCAC4-404466245&15 59 44.59&-21 55 25.1&11.975&11.01&7.644&15.461256&17.393635&14.093049\\
		RIK-56&16 01 13.99&-25 16 28.2&11.357&10.419&6.805&14.797748&15.560816&13.121471\\
		RIK-58&16 01 29.03&-25 09 06.9&11.186&10.084&5.341& & & \\
		UCAC4-4552391&16 02 05.18&-23 31 06.9&11.73&10.673&5.962&15.057345&16.919445&13.7677355\\
		RIK-65&16 02 35.89&-23 20 17.1&11.163&10.202&7.406&14.319887&15.939925&13.069498\\
		UCAC4-423900944&16 03 30.72&-40 24 34.0&11.027&10.016&6.393&14.096236&15.72044&12.857787\\
		RIK-77&16 04 18.93&-24 30 39.3&9.975&8.853&4.522& & & \\
		UCAC4-1264898768&16 05 06.46&-17 34 02.1&11.755&10.785&5.4&15.192967&16.892452&13.807232\\
		RIK-78&16 05 21.57&-18 21 41.2&9.263&8.134&3.156&11.744559&12.398394&10.756141\\
		UCAC4-39082833&16 05 44.04&-34 37 59.0&10.572&9.721&7.28&12.630634&13.44426&11.752831\\
		UCAC4-415825814&16 06 23.82&-18 07 18.4&11.441&10.38&5.287&14.609817&16.326954&13.345088\\
		RIK-80&16 06 43.86&-19 08 05.5&10.137&9.195&6.79&12.624823&13.601917&11.626397\\
		RIK-81&16 06 47.94&-18 41 43.8&9.886&8.979&3.927&12.061529&12.911606&11.1612835\\
		UCAC4-39156047&16 07 08.52&-32 01 01.1&11.22&10.141&5.047&13.704388&14.950171&12.603363\\
		RIK-82&16 07 14.02&-17 02 42.7&11.754&10.754&6.471&14.857619&16.504723&13.609353\\
		UCAC4-853813426&16 07 52.31&-38 58 06.1&11.012&10.009&5.945&14.028573&15.560053&12.817073\\
		UCAC4-853849672&16 08 30.70&-38 28 26.8&8.974&8.225&3.07&10.670469&11.257574&9.964868\\
		RIK-96&16 09 31.66&-22 29 22.4&10.143&9.148&4.228&13.07967&14.475168&11.903089\\
		RIK-102&16 10 05.02&-21 32 31.9&10.069&8.951&3.643&12.503063&13.434556&11.528659\\
		UCAC4-415948957&16 10 14.46&-19 51 37.5&11.214&10.234&7.947&14.400462&16.073004&13.143824\\
		UCAC4-415950015&16 10 19.04&-21 24 25.0&11.743&10.738&5.823&15.097035&17.00309&13.775255\\
		UCAC4-415970246&16 10 43.92&-20 32 02.6&11.401&10.204&6.596&15.141113&16.810104&13.581232\\
		RIK-111&16 11 26.03&-26 31 55.9&10.563&9.568&5.98&12.988873&14.228143&11.845519\\
		RIK-112&16 12 05.05&-20 43 40.5&10.102&9.065&5.931&12.580939&13.55433&11.591411\\
		RIK-113&16 12 06.68&-30 10 27.1&10.35&9.317&3.604&14.028227&15.49386&12.829292\\
		RIK-124&16 13 21.91&-21 36 13.7&11.014&9.977&5.409& & & \\
		RIK-138&16 14 52.41&-25 13 52.4&11.492&10.454&7.522&15.036407&16.464985&13.45865\\
		RIK-146&16 15 32.20&-20 10 23.7&10.158&8.908&4.571& & & \\
		UCAC4-39632762&16 15 35.05&-34 32 01.4&11.082&10.167&6.508&13.573205&14.608051&12.555316\\
		UCAC4-416216449&16 16 46.89&-20 33 32.4&11.585&10.511&5.921&14.955278&16.765795&13.657503\\
		UCAC4-416217728&16 16 50.82&-20 09 08.2&10.604&9.429&5.183&13.776228&15.260837&12.5272455\\
		UCAC4-416239730&16 17 18.90&-22 30 01.7&11.327&10.309&5.638& & & \\
		UCAC4-39740614&16 17 49.05&-32 55 14.6&10.107&9.023&4.551&13.331557&14.181413&11.911238\\
		UCAC4-1056352275&16 18 06.62&-41 26 32.5&10.67&9.666&6.273& & & \\
		\hline
	\end{tabular}
\label{table:magnitudes1}
\end{table*}
\begin{table*}
	\centering
	\begin{tabular}{|c|c|c|c|c|c|c|c|c|}
		Object & RA & DEC & $J$ & $K$ & $W4$ & $G$ & $BP$ & $RP$\\
		\hline
		UCAC4-160914413&16 19 10.09&-24 32 08.9&11.345&10.234&5.559&14.792324&16.690454&13.478447\\		
		RIK-183&16 20 06.16&-22 12 38.5&11.608&10.653&7.204&14.685614&16.286549&13.45464\\
		UCAC4-416359472&16 20 22.91&-22 27 04.1&11.189&10.181&6.632&14.309035&15.80167&13.09051\\
		UCAC4-1253626396&16 21 57.69&-24 29 43.5&8.124&7.587&6.111&9.890488&10.465755&9.178918\\
		UCAC4-1253629107&16 22 01.97&-25 22 20.4&13.334&12.215&6.232& & & \\
		UCAC4-599700591&16 22 11.00&-48 41 38.5&12.53&11.413&6.932& & & \\
		UCAC4-64946760&16 22 39.58&-35 13 06.0&11.882&10.89&6.553&15.00873&16.723364&13.742043\\
		RIK-223&16 25 28.81&-26 07 53.8&10.885&9.854&7.428&13.848455&15.308706&12.659194\\
		RIK-239&16 27 12.74&-25 04 01.8&10.553&9.382&5.483&13.545703&14.904157&12.389707\\
		UCAC4-450968247&16 28 34.97&-20 19 11.1&11.444&10.496&8.223&14.231308&15.337696&13.172604\\
		UCAC4-379613161&16 29 28.82&-25 00 25.0&13.099&11.914&7.927&16.288717&17.511415&15.127983\\
		UCAC4-455511949&16 30 03.71&-42 26 58.7&10.021&8.828&4.243&12.534797&13.566977&11.528833\\
		UCAC4-496722210&16 30 37.96&-29 54 22.4&9.05&8.385&5.71&10.566731&11.138648&9.889597\\
		UCAC4-60335445&16 30 42.12&-26 23 03.9&11.703&10.68&7.804&15.005119&16.632168&13.6854\\
		RIK-250&16 33 34.97&-18 32 53.9&11.314&10.397&7.724&14.444289&16.047361&13.208809\\
		UCAC4-31196367&16 35 11.73&-22 57 28.1&12.127&11.172&5.961&14.711923&15.647959&13.726332\\
		UCAC4-456216791&16 35 27.22&-43 25 27.9&13.843&12.663&6.799& & & \\
		UCAC4-60621830&16 36 46.51&-25 02 03.3&10.062&8.899&4.618&12.824942&14.049942&11.732743\\
		UCAC4-1067529427&16 45 28.96&-25 02 47.7&11.214&10.274&4.95&14.422803&16.055206&13.146103\\
		UCAC4-431964534&16 46 56.03&-32 42 54.0&10.283&9.47&7.202&12.390285&13.218088&11.505731\\
		\hline
	\end{tabular}
	\label{table:magnitudes}
\end{table*}

\begin{table*}
	\centering
	\caption{Full list of objects observed in this survey. P(M) indicates the probability of membership to the listed region, based on Rizzuto et al. (2015) which uses proper motion and Bayesian analysis to determine a probability. An object has is given the label: $N_{Obs}$ lists the number of observations obtained over the course of this survey. Temp. SpT. lists the spectral type of preferred template used to obtain radial velocities (see Section 2.3.3).$<v_r>$ is the mean radial velocity over all epochs. $\Delta v_r$ is the radial velocity variation taken to be the difference between epochs with the highest and lowest radial velocity. The Bayes' factor calculated from the Bayesian analysis described in Section 3.3. is the ratio of likelihoods of being a binary star to being a single star.}
	\begin{tabular}{|c|c|c|c|c|c|c|c|c|c|c|c|}
		Object & Region & RA & DEC & P(M) & Disc? & $N_{Obs}$ & Temp. SpT. & $<v_r>$ & $\Delta v_r$ & Bayes' Factor(binary?)\\
		\hline
		UCAC4-827387234&UCL&13 54 35.61&-43 56 48.1&77&YY&5&M2.5V&11.22&3.52&0.53(N)\\
		UCAC4-467124593&UCL&14 10 59.69&-46 11 33.8&45&YY&5&M2.5V&9.99&4.61&0.52(N)\\
		UCAC4-947158099&UCL&14 14 52.40&-57 22 54.0&72&YY&3&K6Va&-18.43&9.52&1.62(N)\\
		UCAC4-382098779&UCL&14 16 03.10&-40 30 34.2&67&YY&4&M2.5V&5.42&6.06&0.65(N)\\
		UCAC4-945204860&UCL&14 27 30.11&-60 42 18.0&73&YY&2&K3+V&-19.61&1.8&0.59(N)\\
		UCAC4-442012351&UCL&14 34 36.62&-45 45 59.9&53&YY&4&M2.5V&13.02&7.42&0.94(N)\\
		UCAC4-371257082&UCL&14 41 37.66&-37 56 34.4&58&NY&3&M2.5V&6.94&6.98&0.58(N)\\
		UCAC4-371337294&UCL&14 44 15.62&-40 32 27.3&68&YY&3&M2.5V&8.51&2.26&0.54(N)\\
		UCAC4-21594717&UCL&14 50 55.74&-35 04 22.9&25&YY&4&K6Va&5.08&2.39&0.4(N)\\
		UCAC4-365499547&UCL&15 24 35.62&-39 30 43.3&55&YY&4&M2.5V&8.06&4.28&0.61(N)\\
		UCAC4-445278523&UCL&15 29 17.48&-46 36 28.7&69&YY&3&G5V&2.46&70.28&3.87e+89(Y)\\
		UCAC4-23251943&UCL&15 29 23.88&-34 11 54.5&85&YY&4&M2.5V&6.48&2.22&0.52(N)\\
		UCAC4-23288203&UCL&15 30 33.57&-32 46 19.1&57&YY&4&M1.5&8.36&2.82&0.51(N)\\
		UCAC4-447414452&UCL&15 37 22.69&-40 17 59.6&70&YY&5&K0V&6.12&12.95&10101.62(Y)\\
		UCAC4-333669217&US&15 45 09.50&-15 40 29.0&83&YY&4&M2.5V&0.15&6.99&0.58(N)\\
		UCAC4-161328427&UCL&15 46 25.81&-31 43 19.3&41&YY&4&M2.5V&5.2&76.67&2.36e+66(Y)\\
		UCAC4-30196323&US&15 48 24.44&-22 35 49.7&93&YY&4&M2.5V&-0.89&3.21&0.53(N)\\
		UCAC4-30231541&US&15 49 19.76&-22 57 29.7&97&YY&4&M2.5V&0.25&4.65&0.5(N)\\
		UCAC4-1297314504&UCL&15 49 30.74&-35 49 51.4&84&YY&4&M2.5V&3.14&4.34&0.5(N)\\
		UCAC4-408863727&US&15 52 08.84&-27 23 45.8&87&YY&4&M2.5V&4.8&8.5&0.6(N)\\
		UCAC4-1019544336&UCL&15 52 13.44&-39 56 08.2&81&YY&4&M2.5V&8.39&1.17&0.51(N)\\
		RIK-23&US&15 56 42.45&-20 39 34.2&98&YY&5&M2.5V&-2.24&3.76&0.44(N)\\
		RIK-30&US&15 57 34.31&-23 21 12.3&-1&NY&4&M1.5&-0.97&6.83&0.56(N)\\
		UCAC4-519496506&UCL&15 57 43.61&-41 43 37.8&63&YY&4&K6Va&-0.09&2.42&0.34(N)\\
		UCAC4-426579903&US&15 57 54.45&-24 50 42.4&85&YY&5&K3.5V&-5.24&4.73&0.37(N)\\
		UCAC4-404438620&US&15 57 58.92&-18 14 59.6&91&YY&4&K6Va&-4.24&1.88&0.3(N)\\
		RIK-34&US&15 58 12.70&-23 28 36.4&82&NY&6&G7.5IV-V&-4.86&1.72&0.15(N)\\
		UCAC4-519508031&UCL&15 58 20.40&-40 33 05.8&38&YY&5&M2.5V&3.38&2.14&0.42(N)\\
		RIK-38&US&15 58 36.20&-19 46 13.5&95&YY&4&M2.5V&0.02&5.36&0.49(N)\\
		RIK-43&US&15 59 44.27&-20 29 23.4&88&YY&7&M2.5V&-3.14&9.08&0.54(N)\\
		UCAC4-404466245&US&15 59 44.59&-21 55 25.1&90&YY&2&M2.5V&-2.55&2.15&0.62(N)\\
		RIK-56&US&16 01 13.99&-25 16 28.2&-1&YY&6&M2.5V&1.27&2.76&0.41(N)\\
		RIK-58&US&16 01 29.03&-25 09 06.9&-1&YY&6&M2.5V&-2.8&3.58&0.41(N)\\
		UCAC4-4552391&US&16 02 05.18&-23 31 06.9&97&YY&4&M2.5V&0.18&3.56&0.53(N)\\
		RIK-65&US&16 02 35.89&-23 20 17.1&94&YY&4&M2.5V&-1.42&0.39&0.47(N)\\
		UCAC4-423900944&UCL&16 03 30.72&-40 24 34.0&29&YY&5&M2.5V&5.6&1.83&0.45(N)\\
		RIK-77&US&16 04 18.93&-24 30 39.3&-1&YY&4&M2.5V&-2.42&5.28&0.5(N)\\
		UCAC4-1264898768&US&16 05 06.46&-17 34 02.1&94&YY&3&M2.5V&0.91&5.15&0.64(N)\\
		RIK-78&US&16 05 21.57&-18 21 41.2&79&YY&4&K6Va&-3.99&2.51&0.32(N)\\
		UCAC4-39082833&UCL&16 05 44.04&-34 37 59.0&51&YY&4&K6Va&3.87&1.16&0.37(N)\\
		UCAC4-415825814&US&16 06 23.82&-18 07 18.4&86&YY&4&M2.5V&-2.76&6.54&0.53(N)\\
		RIK-80&US&16 06 43.86&-19 08 05.5&85&YY&5&K6Va&-4.83&2.53&0.27(N)\\
		RIK-81&US&16 06 47.94&-18 41 43.8&72&YY&5&K6Va&-3.24&4.21&0.45(N)\\
		UCAC4-39156047&US&16 07 08.52&-32 01 01.1&87&YY&4&M1.5&5.32&8.88&0.86(N)\\
		RIK-82&US&16 07 14.02&-17 02 42.7&97&YY&4&M2.5V&-0.42&4.77&0.52(N)\\
		UCAC4-853813426&UCL&16 07 52.31&-38 58 06.1&74&YY&4&M2.5V&2.56&3.99&0.46(N)\\
		UCAC4-853849672&UCL&16 08 30.70&-38 28 26.8&29&NY&4&K3.5V&1.58&0.88&0.29(N)\\
		RIK-96&US&16 09 31.66&-22 29 22.4&92&YY&6&M1.5&14.62&42.37&7.84e+50(Y)\\
		RIK-102&US&16 10 05.02&-21 32 31.9&91&YY&4&K6Va&-5.43&1.92&0.38(N)\\
		UCAC4-415948957&US&16 10 14.46&-19 51 37.5&78&NY&3&M2.5V&-0.14&2.97&0.56(N)\\
		UCAC4-415950015&US&16 10 19.04&-21 24 25.0&92&YY&3&M2.5V&-4.3&1.87&0.65(N)\\
		UCAC4-415970246&US&16 10 43.92&-20 32 02.6&83&YY&2&M2.5V&0.16&2.18&0.71(N)\\
		RIK-111&US&16 11 26.03&-26 31 55.9&96&YY&4&M2.5V&0.8&1.75&0.43(N)\\
		RIK-112&US&16 12 05.05&-20 43 40.5&95&YY&4&M1.5&-1.87&4.36&0.46(N)\\
		RIK-113&US&16 12 06.68&-30 10 27.1&73&YY&5&M1.5&1.75&2.64&0.41(N)\\
		RIK-124&US&16 13 21.91&-21 36 13.7&93&YY&4&M2.5V&-0.61&2.41&0.49(N)\\
		RIK-138&US&16 14 52.41&-25 13 52.4&97&YY&3&M2.5V&-0.36&2.58&0.55(N)\\
		RIK-146&US&16 15 32.20&-20 10 23.7&-1&YY&7&M2.5V&-2.3&13.12&0.39(N)\\
		\hline
	\end{tabular}
	\label{table:targets1}
\end{table*}

\begin{table*}
	\centering
		\begin{tabular}{|c|c|c|c|c|c|c|c|c|c|c|c|}
		Object & Region & RA & DEC & P(M) & Disc? & $N_{Obs}$ & Temp. SpT. & $<v_r>$ & $\Delta v_r$ & Bayes' Factor(binary?)\\
		\hline
		UCAC4-39632762&US&16 15 35.05&-34 32 01.4&83&NY&4&K6Va&5.55&2.53&0.42(N)\\
		UCAC4-416216449&US&16 16 46.89&-20 33 32.4&97&YY&3&M2.5V&-0.03&3.45&0.63(N)\\
		UCAC4-416217728&US&16 16 50.82&-20 09 08.2&73&YY&3&M2.5V&-2.79&1.62&0.63(N)\\
		UCAC4-416239730&US&16 17 18.90&-22 30 01.7&94&YY&3&M2.5V&-2.89&4.81&0.66(N)\\
		UCAC4-39740614&US&16 17 49.05&-32 55 14.6&94&YY&5&M1.5&4.37&9.87&3.84(N)\\
		UCAC4-1056352275&UCL&16 18 06.62&-41 26 32.5&82&YY&4&M2.5V&3.02&3.51&0.48(N)\\
		UCAC4-160914413&US&16 19 10.09&-24 32 08.9&95&YY&3&M2.5V&2.05&4.09&0.6(N)\\
		RIK-183&US&16 20 06.16&-22 12 38.5&96&YY&4&M2.5V&1.25&3.88&0.46(N)\\
		UCAC4-416359472&US&16 20 22.91&-22 27 04.1&91&YY&3&M2.5V&-2.72&8.62&0.96(N)\\
		UCAC4-1253626396&US&16 21 57.69&-24 29 43.5&80&NY&5&M1V&11.13&26.3&462000000000.0(Y)\\
		UCAC4-1253629107&US&16 22 01.97&-25 22 20.4&57&YY&3&K0V&0.67&15.0&6.31(N)\\
		UCAC4-599700591&UCL&16 22 11.00&-48 41 38.5&77&YY&2&K3+V&-57.93&9.28&1.42(N)\\
		UCAC4-64946760&US&16 22 39.58&-35 13 06.0&88&YY&2&M2.5V&3.32&2.1&0.62(N)\\
		RIK-223&US&16 25 28.81&-26 07 53.8&97&NY&3&M2.5V&0.37&1.22&0.51(N)\\
		RIK-239&US&16 27 12.74&-25 04 01.8&-1&YY&3&M1.5&-1.41&1.15&0.49(N)\\
		UCAC4-450968247&US&16 28 34.97&-20 19 11.1&92&NY&3&G7.5IV-V&-22.0&43.28&3.18e+42(Y)\\
		UCAC4-379613161&US&16 29 28.82&-25 00 25.0&86&YY&3&K3V&22.72&4.87&0.58(N)\\
		UCAC4-455511949&UCL&16 30 03.71&-42 26 58.7&59&YY&3&K6Va&-0.83&2.06&0.37(N)\\
		UCAC4-496722210&US&16 30 37.96&-29 54 22.4&30&YY&3&K3+V&-1.53&1.48&0.36(N)\\
		UCAC4-60335445&US&16 30 42.12&-26 23 03.9&97&YY&2&M2.5V&0.62&1.32&0.67(N)\\
		RIK-250&US&16 33 34.97&-18 32 53.9&96&YY&2&M2.5V&-1.17&4.39&0.7(N)\\
		UCAC4-31196367&US&16 35 11.73&-22 57 28.1&63&YY&3&G5V&3.1&5.77&1.3(N)\\
		UCAC4-456216791&UCL&16 35 27.22&-43 25 27.9&56&YY&3&K6Va&-26.65&5.66&0.56(N)\\
		UCAC4-60621830&US&16 36 46.51&-25 02 03.3&89&YY&4&M1.5&-1.28&3.15&0.48(N)\\
		UCAC4-1067529427&US&16 45 28.96&-25 02 47.7&94&YY&3&M2.5V&-3.29&3.45&0.63(N)\\
		UCAC4-431964534&UCL&16 46 56.03&-32 42 54.0&61&YY&3&K6Va&0.13&1.57&0.53(N)\\
		\hline
	\end{tabular}
	\label{table:targets2}
\end{table*}

\end{document}